# Ice Giant Systems:
## The Scientific Potential of Orbital Missions to Uranus and Neptune

**Leigh N. Fletcher** (School of Physics and Astronomy, University of Leicester, University Road, Leicester, LE1 7RH, UK; leigh.fletcher@le.ac.uk; Tel: +441162523585), **Ravit Helled** (University of Zurich, Switzerland), **Elias Roussos** (Max Planck Institute for Solar System Research, Germany), **Geraint Jones** (Mullard Space Science Laboratory, UK), **Sébastien Charnoz (**Institut de Physique du Globe de Paris, France), **Nicolas André** (Institut de Recherche en Astrophysique et Planétologie, Toulouse, France), **David Andrews** (Swedish Institute of Space Physics, Sweden), **Michele Bannister** (Queens University Belfast, UK), **Emma Bunce** (University of Leicester, UK), **Thibault Cavalié** (Laboratoire d'Astrophysique de Bordeaux, Univ. Bordeaux, CNRS, Pessac, France; LESIA, Observatoire de Paris, Université PSL, CNRS, Sorbonne Université, Univ. Paris Diderot, Sorbonne Paris Cité, Meudon, France), **Francesca Ferri** (Università degli Studi di Padova, Italy), **Jonathan Fortney** (University of Santa-Cruz, USA), **Davide Grassi** (Istituto Nazionale di AstroFisica – Istituto di Astrofisica e Planetologia Spaziali (INAF-IAPS), Rome, Italy), **Léa Griton** (Institut de Recherche en Astrophysique et Planétologie, CNRS, Toulouse, France), **Paul Hartogh** (Max-Planck-Institut für Sonnensystemforschung, Germany), **Ricardo Hueso** (Escuela de Ingeniería de Bilbao, UPV/EHU, Bilbao, Spain), **Yohai Kaspi** (Weizmann Institute of Science, Israel), **Laurent Lamy** (Observatoire de Paris, France), **Adam Masters** (Imperial College London, UK), **Henrik Melin** (University of Leicester, UK), **Julianne Moses** (Space Science Institute, USA), **Oliver Mousis (**Aix Marseille Univ, CNRS, CNES, LAM, Marseille, France), **Nadine Nettleman** (University of Rostock, Germany), **Christina Plainaki** (ASI - Italian Space Agency, Italy), **Jürgen Schmidt (**University of Oulu, Finland), **Amy Simon** (NASA Goddard Space Flight Center, USA), **Gabriel Tobie** (Université de Nantes, France), **Paolo Tortora** (Università di Bologna, Italy), **Federico Tosi** (Istituto Nazionale di AstroFisica – Istituto di Astrofisica e Planetologia Spaziali (INAF-IAPS), Rome, Italy), **Diego Turrini** (Istituto Nazionale di AstroFisica – Istituto di Astrofisica e Planetologia Spaziali (INAF-IAPS), Rome, Italy)

**Abstract:**
Uranus and Neptune, and their diverse satellite and ring systems, represent the least explored environments of our Solar System, and yet may provide the archetype for the most common outcome of planetary formation throughout our galaxy. Ice Giants will be the last remaining class of Solar System planet to have a dedicated orbital explorer, and international efforts are under way to realise such an ambitious mission in the coming decades. In 2019, the European Space Agency released a call for scientific themes for its strategic science planning process for the 2030s and 2040s, known as *Voyage 2050*. We used this opportunity to review our present-day knowledge of the Uranus and Neptune systems, producing a revised and updated set of scientific questions and motivations for their exploration. This review article describes how such a mission could explore their origins, ice-rich interiors, dynamic atmospheres, unique magnetospheres, and myriad icy satellites, to address questions at the heart of modern planetary science. These two worlds are superb examples of how planets with shared origins can exhibit remarkably different evolutionary paths: Neptune as the archetype for Ice Giants, whereas Uranus may be atypical. Exploring Uranus' natural satellites and Neptune's captured moon Triton could reveal how Ocean Worlds form and remain active, redefining the extent of the habitable zone in our Solar System. For these reasons and more, we advocate that an Ice Giant System explorer should become a strategic cornerstone mission within ESA's Voyage 2050 programme, in partnership with international collaborators, and targeting launch opportunities in the early 2030s.

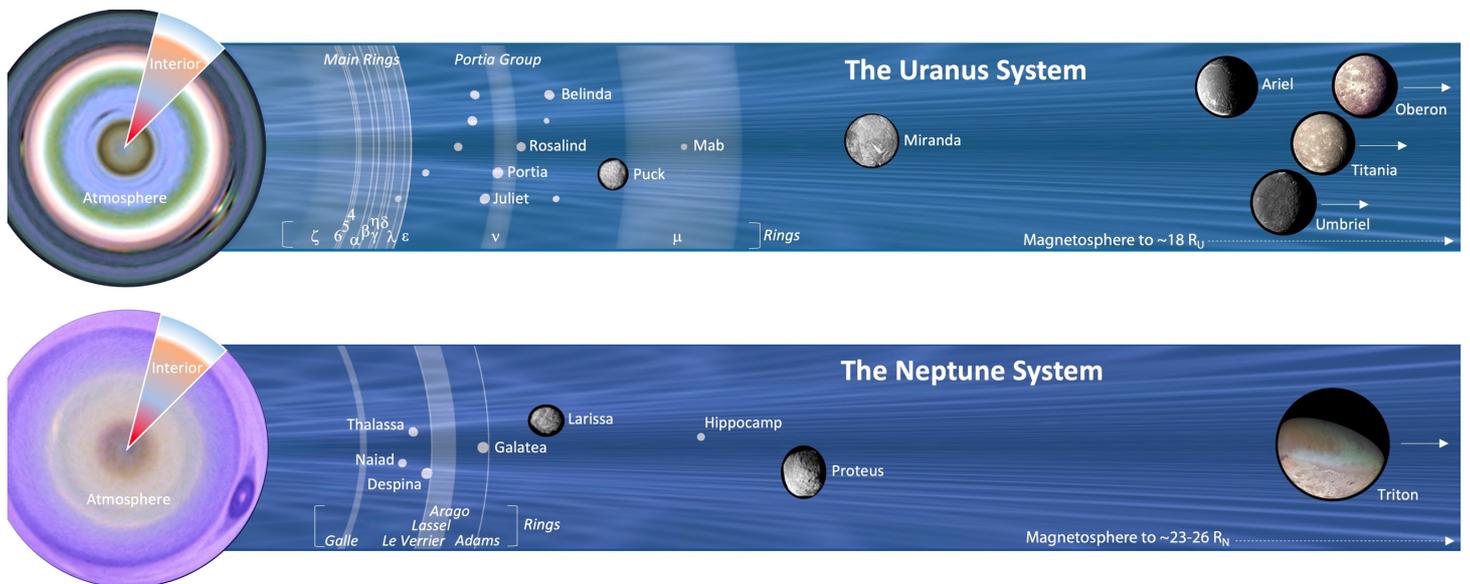

*Figure 1 Each Ice Giant exhibits a rich system of planetary environments to explore, from their mysterious interiors, atmospheres and magnetospheres, to the diverse satellites and rings. The inner systems are to scale, with arrows next to major moons indicating that they orbit at larger planetocentric distances. The magnetosphere and radiation belts would encompass the full area of the figure. Credit: L.N. Fletcher/M. Hedman/E. Karkoschka/Voyager-2.*

## 1. Introduction: Why Explore Ice Giant Systems?

### 1.1 Motivations

The early 21$^{st}$ century has provided unprecedented leaps in our exploration of the Gas Giant systems, via the completion of the Galileo and Cassini orbital missions at Jupiter and Saturn; NASA/Juno's ongoing exploration of Jupiter's deep interior, atmosphere, and magnetic field; ESA's development of the JUICE mission (Jupiter Icy Moons Explorer) and NASA's Europa Clipper to explore the Galilean satellites. The past decade has also provided our first glimpses of the diversity of planetary environments in the outer solar system, via the New Horizons mission to Pluto. Conversely, the realm of the Ice Giants, from Uranus (20 AU) to Neptune (30 AU), remains largely unexplored, each system having been visited only once by a flyby spacecraft – Voyager 2 – in 1986 and 1989 respectively. More than three decades have passed since our first close-up glimpses of these worlds, with cameras, spectrometers and sensors based on 1960s and 70s technologies. Voyager's systems were not optimised for the Ice Giants, which were considered to be extended mission targets. Uranus and Neptune have therefore never had a dedicated mission, despite the rich and diverse systems displayed in Figure 1. A return to the Ice Giants with an orbiter is the next logical step in humankind's exploration of our Solar System.

The Ice Giants may be our closest and best representatives of a whole class of astrophysical objects, as Neptune-sized worlds have emerged as the dominant category in our expanding census of exoplanets (Fulton et al., 2018), intermediate between the smaller terrestrial worlds and the larger hydrogen-rich gas giants (Section 3.3). Our own Ice Giants offer an opportunity to explore physical and chemical processes within these planetary systems as the archetype for these distant exoplanets (Rymer et al., 2019). Furthermore, the formation and evolution of Uranus and Neptune (Section 2.1) pose a critical test of our understanding of planet formation, and of the architecture of our Solar System. Their small size, compared to Jupiter, places strong constraints on the timing of planet formation. Their bulk internal composition (i.e., the fraction of rock, ices, and gases) and the differentiation with depth (i.e., molecular weight gradients, phase transitions to form global water oceans and icy mantles) are poorly known, but help determine the conditions and dynamics in the outer planetary nebula at the time of planet formation.

Uranus and Neptune also provide two intriguing endmembers for the Ice Giant class. Neptune may be considered the archetype for a seasonal Ice Giant, whereas the cataclysmic collision responsible for Uranus' extreme tilt renders it unique in our Solar System. Contrasting the conditions on these two worlds provides insights into differential evolution from shared origins. The atmospheres of Uranus and Neptune (Section 2.2) exemplify the contrasts between these worlds. Uranus' negligible internal heat renders its atmosphere extremely sluggish, with consequences for storms, meteorology, and atmospheric chemistry. Conversely, Neptune's powerful winds and rapidly-evolving storms demonstrates how internal energy can drive powerful weather, despite weak sunlight at 30 AU. Both of these worlds exhibit planetary banding, although the atmospheric circulation responsible for these bands (and their associated winds, temperatures, composition and clouds) remain unclear, and the connection to atmospheric flows below the topmost clouds remains a mystery.

Conditions within the Ice Giant magnetospheres (Section 2.3) are unlike those found anywhere else, with substantial offsets between their magnetic dipole axes and the planets' rotational axes implying a system with an extremely unusual interaction with the solar wind and internal plasma processes, varying on both rotational cycles as the planet spins, and on orbital cycles.

The diverse Ice Giant satellites (Section 2.4) and narrow, incomplete ring systems (Section 2.5) provide an intriguing counterpoint to the better-studied Jovian and Saturnian systems. Uranus may feature a natural, primordial satellite system with evidence of extreme and violent events (e.g., Miranda). Neptune hosts a captured Kuiper Belt Object, Triton, which may itself harbour a sub-surface ocean giving rise to active surface geology (e.g., south polar plumes and cryovolcanism).

Advancing our knowledge of the Ice Giants and their diverse satellite systems requires humankind's first dedicated explorer for this distant realm. Such a spacecraft should combine interior science via gravity and magnetic measurements, *in situ* measurements of their plasma and magnetic field environments, *in situ* sampling of their chemical composition, and close-proximity multi-wavelength remote sensing of the planets, their rings, and moons. This review article will summarise our present understanding of these worlds, and propose a revised set of scientific questions to guide our preparation for such a mission. This article is motivated by ESA's recent call for scientific themes as part of its strategic space mission planning in the period from 2035 to 2050,[1] and is therefore biased to European perspectives on an Ice Giant mission, as explored in the next section.

## 1.2 Ice Giants in ESA's Cosmic Vision

The exploration of the Ice Giants addresses themes at the heart of ESA's existing Cosmic Vision[2] programme, namely (1) exploring the conditions for planet formation and the emergence of life; (2) understanding how our solar system works; and (3) exploring the fundamental physical laws of the universe. European-led concepts for Ice Giant exploration have been submitted to several ESA Cosmic Vision competitions. The Uranus Pathfinder mission, an orbiting spacecraft based on heritage from Mars Express and Rosetta, was proposed as a medium-class (~€0.5bn) mission in both the M3 (2010) and M4 (2014) competitions (Arridge et al., 2012). However, the long duration of the mission, limited power available, and the programmatic implications of having NASA provide the launch vehicle and radioisotope thermoelectric generators (RTGs), meant that the Pathfinder concept did not proceed to the much-needed Phase A study.

The importance of Ice Giant science was reinforced by multiple submissions to ESA's call for large-class mission themes in 2013: a Uranus orbiter with atmospheric probe (Arridge et al., 2014), an orbiter to explore Neptune and Triton (Masters et al., 2014); and a concept for dual orbiters of both worlds (Turrini et al., 2014). Once again, an ice giant mission failed to proceed to the formal study phase, but ESA's Senior Survey Committee (SSC[3]) commented that ``*the exploration of the icy giants appears to be a timely milestone, fully appropriate for an L class mission. The whole planetology community would be involved in the various aspects of this mission... the SSC recommends that every effort is made to pursue this theme through other means, such as cooperation on missions led by partner agencies.*"

This prioritisation led to collaboration between ESA and NASA in the formation of a science definition team (2016-17), which looked more closely at a number of different mission architectures for a future mission to the Ice Giants (Hofstadter et al., 2019). In addition, ESA's own efforts to develop nuclear power sources for space applications have been progressing, with

---

prototypes now developed to utilise the heat from the decay of [241]Am as their power source (see Section 4.3), provided that the challenge of their low energy density can be overcome. Such an advance might make smaller, European-led missions to the Ice Giants more realistic, and addresses many of the challenges faced by the original Uranus Pathfinder concepts.

At the start of the 2020s, NASA and ESA are continuing to explore the potential for an international mission to the Ice Giants. A palette of potential contributions (M-class in scale) to a US-led mission have been identified by ESA[4], and US scientists are currently undertaking detailed design and costing exercises for missions to be assessed in the upcoming US Planetary Decadal Survey (~2022). Each of these emphasise launch opportunities in the early 2030s (Section 4.2), with arrival in the early 2040s (timelines for Ice Giant missions will be described in Section 4.3). This review article summarises those studies, whilst taking the opportunity to update and thoroughly revise the scientific rationale for Ice Giant missions compared to Arridge et al. (2012, 2014), Masters et al. (2014), Turrini et al. (2014) and Hofstadter et al. (2019). We focus on the science achievable from orbit, as the science potential of in situ entry probes has been discussed elsewhere (Mousis et al., 2018). Section 2 reviews our present-day knowledge of Ice Giant Systems; Section 3 places the Ice Giants into their wider exoplanetary context; Section 4 briefly reviews the recent mission concept studies and outstanding technological challenges; and Section 5 summarises our scientific goals at the Ice Giants at the start of the 2020s.

## 2. Science Themes for Ice Giant Exploration

In this section we explore the five multi-disciplinary scientific themes that could be accomplished via orbital exploration of the Ice Giants, and show how in-depth studies of fundamental processes at Uranus and Neptune would have far-reaching implications in our Solar System and beyond. Each sub-section is organised via a series of high-level questions that could form the basis of a mission traceability matrix.

### 2.1 Ice Giant Origins and Interiors

What does the origin, structure, and composition of the two Ice Giants reveal about the formation of planetary systems? Understanding the origins and internal structures of Uranus and Neptune will substantially enhance our understanding of our own Solar System and intermediate-mass exoplanets. Their bulk composition provides crucial constraints on the conditions in the solar nebula during planetary formation and evolution.

***How did the Ice Giants first form, and what constraints can be placed on the mechanisms for planetary accretion?*** The formation of Uranus and Neptune has been a long-standing problem for planet formation theory (e.g., Pollack et al., 1996, Dodson-Robinson & Bodenheimer, 2010, Helled & Bodenheimer, 2014). Yet, the large number of detected exoplanets with sizes comparable (or smaller) to that of Uranus and Neptune suggests that such planetary objects are very common (e.g., Batalha et al. 2013), a fact that is in conflict with theoretical calculations.

The challenge for formation models is to prevent Uranus and Neptune from accreting large amounts of hydrogen-helium (H-He) gas, like the Gas Giants Jupiter and Saturn, to provide the correct final mass and gas-to-solids ratios as inferred by structure models. In the standard planet formation model, core accretion (see Helled et al., 2014 for review and the references therein), a slow planetary growth is expected to occur at large radial distances where the solid surface density is lower, and the accretion rate (of planetesimals) is significantly smaller. For the current locations of Uranus and Neptune, the formation timescale can be comparable to the lifetimes of protoplanetary disks. Due to long accretion times at large radial distances, the formation process is too slow to reach the phase of runaway gas accretion, before the gas disk disappears, leaving behind an intermediate-mass planet (a failed giant planet), which consists mostly of heavy elements and a small fraction of H-He gas.

However, since the total mass of H-He in both Uranus and Neptune is estimated to be 2-3 Earth masses ($M_\oplus$), it implies that gas accretion had already begun, and this requires that the gas disk disappears at a very specific time, to prevent further gas accretion onto the planets. This is known as the ***fine-tuning*** problem in Uranus/Neptune formation (e.g., Venturini & Helled, 2017). Another possibility is that Uranus and Neptune formed *in situ* within a few Mys by pebble accretion. In this formation scenario, the core's growth is more efficient than in the planetesimal accretion case, and the pebble isolation mass is above 20 M⊕. As a result, the forming planets could be heavy-element dominated with H-He envelopes that are metal-rich due to sublimation of icy pebbles (e.g., Lambrechts et al., 2014).

Measuring the elemental abundances in the atmospheres of Uranus and Neptune can provide information on their formation history by setting limits on their formation locations and/or the type of solids (pebbles/planetesimals) that were accreted. Measurements of the elemental abundances of well-mixed noble gases, which are only accessible via *in situ* entry probes, would be particularly informative (e.g. Mousis et al., 2018). In addition, determining the atmospheric metallicity provides valuable constraints for structure models, as discussed below.

***What is the role of giant impacts in explaining the differences between Uranus and Neptune?*** Uranus and Neptune are somewhat similar in terms of mass and radius, but they also have significant differences such as their tilt, internal heat flux, and satellite system. It is possible that these observed differences are a result of giant impacts that occurred after the formation of the planets (e.g., Safronov 1966, Stevenson 1986). An oblique impact of a massive impactor can explain Uranus' spin and lead to the formation of a disk where the regular satellites form. Neptune could have also suffered a head-on impact that could have reached the deep interior, providing sufficient energy (and mass) to explain the higher heat flux, and possibly the higher mass and moment of inertia value (e.g., Stevenson 1986, Podolak & Helled, 2012). Giant impact simulations by various groups confirmed that Uranus' tilt and rotation could be explained by a giant impact (Slattery et al., 1992, Kegerreis et al., 2019). Nevertheless, alternative explanations have been proposed: Boue and Laskar (2010) showed that under special circumstances, the high obliquity could potentially arise from orbital migration without the need for a collision. Understanding the cause of Uranus' tilt and the mechanisms that led to the observed differences between the planets are key questions in planetary science.

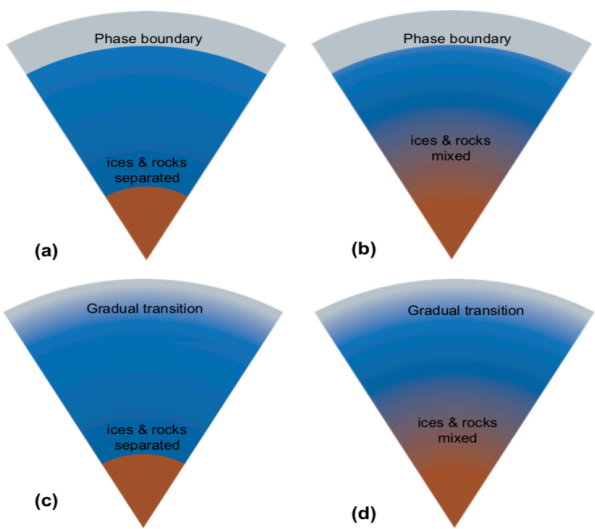

*Figure 2 Sketches of the possible internal structures of the ice giants. It is unclear whether the planets are differentiated and whether the transitions between the different layers are distinct or gradual: (a) separation between the ices and rocks and the ice and H-He atmosphere and a gradual transition between ice and rock, (b) separation (phase boundary) between the H-He atmosphere and ices and a gradual transition between ice and rock, (c) gradual transition between the H-He atmosphere and ice layer, and a distinct separation between the ice and rock layers, and (d) gradual transition both between the H-He atmosphere and ice and the ice and rocks suggesting a global composition gradient with the planets (see text for discussion).*

***What is the bulk composition and internal structure of Uranus and Neptune?*** There are still substantial uncertainties regarding the bulk compositions and internal structures of Uranus and Neptune (e.g., Podolak et al., 1995, Podolak & Helled, 2012, Nettelmann et al., 2013). The available measurements of their physical properties such as mass, radius, rotation rate, atmospheric temperature, and gravitational and magnetic fields are used to constrain models of their interiors. For the Ice Giants, only the gravitational harmonic coefficients $J_2$ and $J_4$ are known and their error bars are relatively large (Jacobson, 2009; 2014), nevertheless, various studies have aimed to constrain the planets' internal structures.

Standard structure models of the planets consist of three layers: a rocky core, an 'icy' shell (water, ammonia, methane, etc.), and a gaseous envelope composed of H-He and heavier elements. The middle layer is not made of "ice" in regard to the physical state of the material (i.e., solid), but is referred to as an icy layer since it is composed of volatile materials such as water, ammonia and methane. The masses and compositions of the layers are modified until the model fits the observed parameters using a physical equation of state (EOS) to represent the materials. Three-layer models predict very high ice-to-rock ratios, where the ice fraction is found to be higher than the rock fraction by 19-35 times for Uranus, and 4-15 times for Neptune, with the total H-He mass typically being 2 and 3 $M_\oplus$ for Uranus and Neptune, respectively. The exact estimates are highly model-dependent, and are sensitive to the assumed composition, thermal structure and rotation rate of the planets (Helled et al., 2011, Nettelmann et al., 2013).

The interiors of Uranus and Neptune could also be more complex with the different elements being mixed, and could also include thermal boundary layers and composition gradients. Indeed, alternative structure models of the planets suggested that Uranus and Neptune could have a density profile without discontinuities (e.g., Helled et al., 2011), and that the planets do not need to contain large fractions of water to fit their observed properties. This alternative model implies that Uranus and Neptune may not be as water-rich as typically thought, but instead are rock-dominated like Pluto (e.g., McKinnon et al., 2017) and could be dominated by composition gradients. It is therefore possible that the "ice giants" are in fact not ice-dominated (see Helled et al., 2011 for details). The very large ice-to-rock ratios found from structure models also suggest a more complex interior structure. At the moment, we have no way to discriminate among the different ice-to-rock ratios inferred from structure models. As a result, further constraints on the gravity and magnetic data (Section 2.3), as well as atmospheric composition and isotopic ratios (e.g., D/H, Atreya et al., 2020) are required.

***How can Ice Giant observations be used to explore the states of matter (e.g., water) and mixtures (e.g., rocks, water, H-He) under the extreme conditions of planetary interiors?*** In order to predict the mixing within the planets, knowledge from equation-of-state (EOS) calculations is required. Internal structure models must be consistent with the phase diagram of the assumed materials and their mixtures. This is a challenging task and progress in that direction is ongoing. EOS calculations can be used to guide model assumptions. For example, it is possible that Uranus and Neptune have deep water oceans that begin where $H_2$ and $H_2O$ become insoluble (e.g., Bailey & Stevenson, 2015, Bali et al., 2013). Figure 2 presents sketches of four possible internal structures of the ice giants where the transitions between layers are distinct (via phase/thermal boundaries) and/or gradual.

Current observational constraints, foremost $J_2$ and $J_4$,

clearly indicate that the deep interior is more enriched in heavy elements than the outer part. Understanding the nature and origin of the compositional gradient zone would yield important information on the formation process and subsequent evolution including possible processes such as outgassing, immiscibility, and sedimentation of ices; processes that play a major role on terrestrial planets and their habitability.

***What physical and chemical processes during the planetary formation and evolution shape the magnetic field, thermal profile, and other observable quantities?*** Structure models must be consistent with the observed multi-polar magnetic fields (see Section 2.3), implying that the outermost ~20% of the planets is convective and consists of conducting materials (e.g., Stanley & Bloxham, 2004, 2006). Currently, the best candidate for the generation of the dynamo is the existence of partially dissociated fluid water in the outermost layers (e.g., Redmer et al.,2011), located above solid and non-convecting superionic water ice 'mantle' (Millot et al., 2019). Dynamo models that fit the Voyager magnetic field data suggest the deep interior is stably stratified (Stanley & Bloxham 2004, 2006) or, alternatively, in a state of thermal-buoyancy driven turbulent convection (Soderlund et al., 2013). Improved measurements of the magnetic fields of Uranus and Neptune will also help to constrain the planetary rotation rate. Since Voyager's measurements of the periodicities in the radio emissions and magnetic fields have not been confirmed by another spacecraft, it is unclear whether the Voyager rotation rate reflects the rotation of the deep interior (Helled et al., 2010), with major consequences for the inferred planetary structure and the question of similar or dissimilar interiors (Nettelmann et al., 2013).

Finally, the different intrinsic heat fluxes of Uranus and Neptune imply that they followed different evolutionary histories. This could be a result of a different growth history or a result of giant impacts during their early evolution (e.g., Reinhardt et al., 2020). Moreover, thermal evolution models that rely on Voyager's measurements of the albedo, brightness temperatures, and atmospheric pressure-temperature profiles that are used to model the evolution of the two planets cannot explain both planets with the same set of assumptions.

**Summary:** A better understanding of the origin, evolution and structure of Ice Giant planets requires new and precise observational constraints on the planets' gravity field, rotation rate, magnetic field, atmospheric composition, and atmospheric thermal structure, both from orbital observations and in situ sampling from an atmospheric probe. The insights into origins, structures, dynamo operation and bulk composition provided by an Ice Giant mission would not only shed light on the planet-forming processes at work in our Solar System, but could also help to explain the most common planetary class throughout our observable universe.

## 2.2 Ice Giant Atmospheres

Why do atmospheric processes differ between Uranus, Neptune, and the Gas Giants, and what are the implications for Neptune-mass worlds throughout our universe? Ice Giant atmospheres occupy a wholly different region of parameter space compared to their Gas Giant cousins. Their dynamics and chemistry are driven by extremes of internal energy (negligible on Uranus, but powerful on Neptune) and extremes of solar insolation (most severe on Uranus due to its 98° axial tilt) that are not seen anywhere else in the Solar System (Figure 3). Their smaller planetary radii, compared to Jupiter and Saturn, affects the width of zonal bands and the drift behaviour of storms and vortices. Their zonal winds are dominated by broad retrograde equatorial jets and do not exhibit the fine-scale banding found on Jupiter, which means that features like bright storms and dark vortices are able to drift with latitude during their lifetimes. Their hydrogen-helium atmospheres are highly enriched in volatiles like $CH_4$ and $H_2S$ that show strong equator-to-pole gradients, changing the atmospheric density and hence the vertical shear on the winds (Sun et al., 1991). Their temperatures are so low that the energy released from interconversion between different states of hydrogen (ortho and para spin isomers) can play a role in shaping atmospheric dynamics (Smith & Gierasch, 1995). Their middle and upper atmospheres are both much hotter than can be explained by weak solar heating alone, implying a decisive role for additional energy from internal (e.g., waves) or external sources (e.g., currents induced by complex coupling to the magnetic field). As the *atmospheres are the windows through which we interpret the bulk properties of planets*, these defining properties of Ice Giants can provide insights into atmospheric processes on intermediate-sized giant planets beyond our Solar System.

A combination of global multi-wavelength remote sensing from an orbiter and *in situ* measurements from an entry probe would provide a transformative understanding of these unique atmospheres, focussing on the following key questions (summarised in Figure 3):

***What are the dynamical, meteorological, and chemical impacts of the extremes of planetary luminosity?*** Despite the substantial differences in self-luminosity (Pearl et al., 1990, 1991), seasonal influences, atmospheric activity (Hueso and Sanchez-Lavega et al., 2019), and the strength of vertical mixing (resulting in differences in atmospheric chemistry, Moses et al., 2018), there are many similarities between these two worlds. In their upper troposphere, tracking of discrete cloud features has revealed that both planets exhibit broad retrograde jets at their equators and prograde jets nearer the poles (Sanchez-Lavega et al., 2018), but unlike Jupiter, these are seemingly disconnected from the fine-scale banding revealed in the visible and near-infrared (Sromovsky et al., 2015). Is this simply an observational bias, or are winds on the ice giants truly different from those on Jupiter and Saturn? What sets the scales of the bands? On the Gas Giants, small-scale eddies (from atmospheric instabilities and convective storms) appear to feed energy into the large-scale winds, but we have never been able to investigate similar processes on Uranus and Neptune. Indeed, convective processes themselves could be substantially different – moist convection driven by the condensation of water will likely play a very limited role in the observable atmosphere, as water is restricted to pressures that exceed tens or hundreds of bars. Instead, convection in the observable troposphere may be driven by methane condensation in the 0.1-1.5 bar range (Stoker & Toon, 1989), or by heat release by ortho-para-$H_2$ conversion (Smith & Gierasch, 1995). These sources of energy operate in a very different way compared to those available on Jupiter and Saturn. Firstly, the high enrichment in methane in the Ice Giants could inhibit vertical motions due to a vertical gradient of the atmospheric molecular weight (Guillot et al., 1995; Leconte et al., 2017). This phenomenon may also be at work in the deep and inaccessible water clouds of both the Gas and Ice giants, but the observable methane clouds of Uranus and Neptune provide an excellent opportunity to study it (Guillot et al., 2019). Secondly, heat release by ortho-para $H_2$ conversion is much more efficient in providing energy within the cold atmospheres of Uranus and Neptune. Thus, convection may occur in vertically-thin layers (Gierasch et al., 1987), rather than extending vertically over tremendous heights, or in a complex and inhomogeneous weather layer (Hueso et al., 2020). Multi-wavelength remote sensing of the temperatures, clouds, winds, and gaseous composition is required to investigate how these meteorological processes differ from the Gas Giants, how they derive their energy from the internal heating or weak sunlight, and their relation to the large-scale banded patterns and winds (Fletcher et al., 2020). Spatially-resolved reflectivity and thermal-emission mapping will allow precise constraints on the Ice Giant energy balance to constrain their self-luminosities. And mapping the distribution and depth of Ice Giant lightning, previously detected via radio emissions on both worlds (Zarka et al., 1986; Gurnett et al., 1990), could determine the frequency and intensity of water-powered convection on the Ice Giants, elucidating its impact on their atmospheric dynamics.

***What is the large-scale circulation of Ice Giant atmospheres, and how deep does it go?*** Atmospheric circulation, driven by both internal energy and solar heating, controls the thermal structure, radiative energy balance, condensate cloud and photochemical haze characteristics, and meteorology. Unlike Jupiter and Saturn, observations from Voyager, space telescopes, and ground-based observatories have revealed mid-latitude upwelling (where most of the vigorous storms and coolest temperatures are found) and sinking motions at the equator and poles (e.g., Conrath et al., 1998, Fletcher et al., 2020). This is superimposed onto polar depletions in several key cloud-forming volatiles: methane (from reflection spectroscopy, Sromovsky et al., 2014; Karkoschka et al.,

2011); hydrogen sulphide (from near-IR and microwave spectroscopy, de Pater et al., 1991; Hofstadter & Butler, 2003; Irwin et al., 2018); and potentially ammonia (from microwave imaging). Do these contrasts imply circulation patterns extending to great depths (de Pater et al., 2014), or are they restricted in vertically-thin layers (Gierasch et al., 1987)? Recent re-analysis of the gravity fields measured by Voyager (Kaspi et al., 2013) suggests that zonal flows are restricted to the outermost ~1000 km of their radii, indicating relatively shallow weather layers overlying the deep and mysterious water-rich interiors. The circulation of the stratosphere is almost entirely unknown on both worlds, due to the challenge of observing weak emissions from hydrocarbons in the mid-infrared (e.g., Orton et al., 2014; Roman et al. 2020). Either way, observations of Uranus and Neptune will have stark implications for atmospheric circulation on intermediate-sized planets with strong chemical enrichments and latitudinal gradients.

***How does atmospheric chemistry and haze formation respond to extreme variations in sunlight and vertical mixing, and the influence of external material?*** Methane can be transported into the stratosphere, where photolysis initiates rich chemical pathways to produce a plethora of hydrocarbons (Moses et al., 2018). The sluggish mixing of Uranus indicates that its photochemistry occurs in a different physical regime (higher pressures) than on any other world. Furthermore, oxygen compounds from external sources (from cometary impacts, infalling dust, satellite and ring

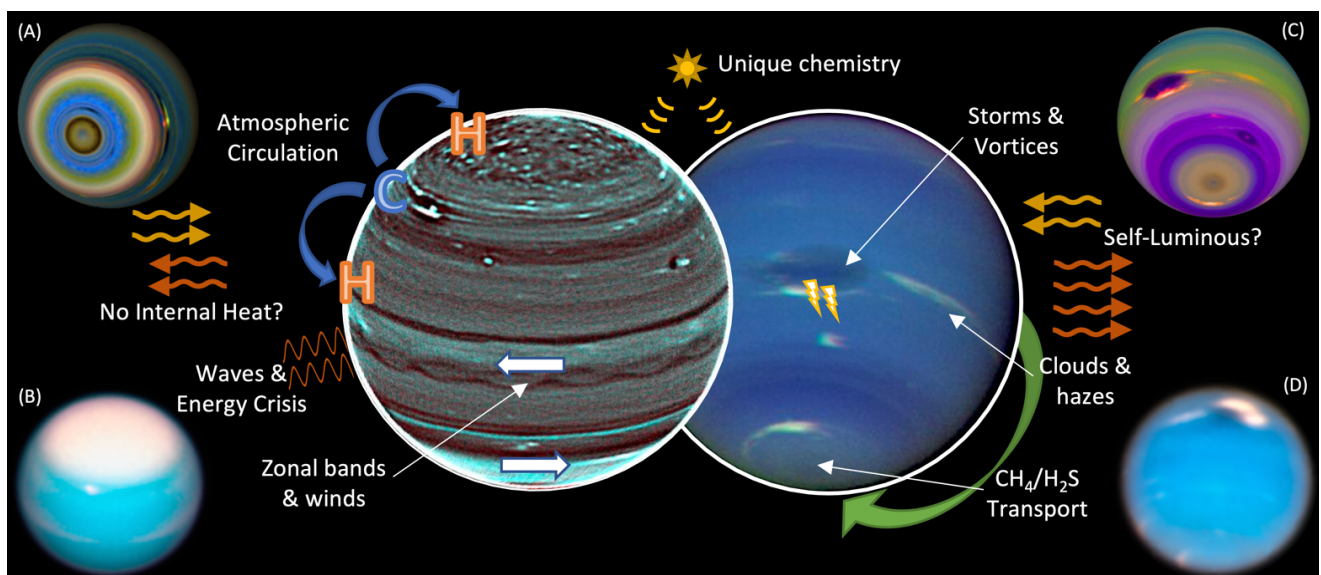

*Figure 3 Ice Giant atmospheres are shaped by dynamical, chemical and radiative processes that are not found elsewhere in our Solar System. Images A & C are false-colour representations of Voyager 2 observations of Uranus and Neptune, respectively. Images B and D were acquired by the Hubble Space Telescope in 2018.*

material, Feuchtgruber et al., 1997) will play different photochemical roles on Uranus, where the methane homopause is lower, than on Neptune (Moses & Poppe, 2017). This exogenic influence can further complicate inferences of planetary formation mechanisms from measurements of bulk abundances (particularly for CO). This rich atmospheric chemistry will be substantially different from that on the Gas Giants, due to the weaker sunlight, the colder temperatures (changing reaction rates and condensation processes, Moses et al., 2018), and the unusual ion-neutral chemistry resulting from the complex magnetic field tilt and auroral processes (Dobrijevic et al. 2020). Condensation of these chemical products (and water ice) can form thin haze layers observed in high-phase imaging (Rages et al., 1991), which may add to the radiative warming of the stratosphere, be modulated by vertically-propagating waves, and could sediment downwards to serve as condensation nuclei or aerosol coatings in the troposphere. Furthermore, Uranus' axial tilt presents an extreme test of coupled chemical and transport models, given that each pole can spend decades in the shroud of winter darkness. The strength of vertical mixing may vary with location, and disequilibrium tracers can be used to assess where mixing is strongest (Fletcher et al., 2020; Cavalie et al., 2020). Such tracers include CO (Cavalie et al., 2017), para-$H_2$ (Conrath et al., 1998) and yet-to-be-detected phosphine (Moreno et al., 2009), arsine, germane, silane, or even tropospheric hydrocarbons (ethane, acetylene). Fluorescence spectroscopy, infrared emissions and sub-mm sounding will reveal the vertical, horizontal, and temporal variability of the chemical networks in the unique low-temperature regimes of Uranus and Neptune.

***What are the sources of energy responsible for heating the middle- and upper-atmospheres?*** Weak sunlight alone cannot explain the high temperatures of the stratosphere (Li et al., 2018) and thermosphere (Herbert et al., 1987), and this severe deficit is known as the energy crisis. Exploration of Uranus and Neptune may provide a solution to this problem, potentially revealing how waves transport energy vertically from the convective troposphere into the middle atmosphere, and how the currents induced by the asymmetric and time-variable magnetic fields provide energy to the upper atmosphere via Joule heating. For example, the long-term cooling of Uranus' thermosphere, observed via emission from $H_3^+$ in the ionosphere (Melin et al., 2019), appears to follow Uranus' magnetic season, which may hint at the importance of particle precipitation modulated by the magnetosphere in resolving the energy crisis. Solving this issue at Uranus or Neptune, via wave observations and exploring magnetosphere-ionosphere-atmosphere coupling processes (e.g., via aurora detected in the UV and infrared), will provide insights into the energetics of all planetary atmospheres.

***How do planetary ionospheres enable the energy transfer that couples the atmosphere and magnetosphere?*** In-situ radio occultations remain the only source for the vertical distribution of electron density in the ionosphere (Majeed et al., 2004), a critical parameter for determining the strength of the coupling between the atmosphere and the magnetosphere. The Voyager 2 occultations of both Uranus and Neptune (Lindal et al, 1986, 1992) provided only two profiles for each planet, providing very poor constraints on what drives the complex shape of the electron density profiles in the ionosphere, including the influx of meteoritic material (Moses et al., 2017).

***How do Ice Giant atmospheres change with time?*** In the decades since the Voyager encounters, Uranus has displayed seasonal polar caps of reflective aerosols with changing winds (Sromovsky et al., 2015) and long-term upper atmospheric changes (Melin et al., 2019); Neptune's large dark anticyclones – and their associated orographic clouds – have grown, drifted, and dissipated (Lebeau et al., 1998, Stratman et al., 2001, Wong et al., 2018, Simon et al., 2019); and a warm south polar vortex developed and strengthened during the Neptunian summer (Fletcher et al., 2014). What are the drivers for these atmospheric changes, and how do they compare to the other planets? There have been suggestions that storm activity has occurred episodically, potentially with a seasonal connection (de Pater et al., 2015; Sromovsky et al., 2015), but is this simply driven by observational bias to their sunlit hemispheres? Mission scenarios for the early 2040s would result in observations separated from those obtained by the Voyager 2 by 0.5 Uranian years and 0.25 Neptunian years. Orbital remote sensing over long time periods, sampling both summer and winter hemispheres, could reveal the causes of atmospheric changes in a regime of extremely weak solar forcing, in contrast to Jupiter and Saturn.

**Summary: Investigations of dynamics, chemistry, cloud formation, atmospheric circulation, and energy transport on Uranus and Neptune would sample a sizeable gap in our understanding of planetary atmospheres, in an underexplored regime of weak**

**seasonal sunlight, low temperatures, and extremes of internal energy and vertical mixing.**

## 2.3 Ice Giant Magnetospheres

What can we learn about astrophysical plasma processes by studying the highly-asymmetric, fast-rotating Ice Giant magnetospheres? The off-centered, oblique and fast rotating planetary magnetic fields of Uranus and Neptune give rise to magnetospheres that are governed by large scale asymmetries and rapidly evolving configurations (e.g. Griton et al. 2018), with no other parallels in our Solar System. The solar wind that impinges upon these two magnetospheres attains Mach numbers significantly larger than those found at Earth, Jupiter, and Saturn, adding further to their uniqueness (Masters et al. 2018). Magnetospheric observations should thus be a high priority in the exploration of the Ice Giants because they extend the sampling of the vast parameter space that controls the structure and evolution of magnetospheres, thus allowing us to achieve a more universal understanding of how these systems work. Insights would also be provided to astrophysical plasma processes on similar systems that are remote to us both in space and time. Such may be the magnetospheres of exoplanets or even that of the Earth at times of geomagnetic field reversals, when the higher order moments of the terrestrial magnetic field become significant, as currently seen at the Ice Giants' (Merrill and Mcfadden, 1999). Evidence for $H_3^+$ ionospheric temperature modulations at Uranus due to charged particle precipitation (Melin et al. 2011; 2013) is one of many indications reminding us how strong a coupling between a planet and its magnetosphere can be, and why the study of the latter would be essential also for achieving a system-level understanding of the Ice Giants.

A synergy between close proximity, remote sensing, and in-situ magnetospheric measurements at the Ice Giants would redefine the state-of-the-art, currently determined by the single Voyager-2 flyby measurements, and limited Earth-based auroral observations. Key questions that would guide such observations are listed below:

***Is there an equilibrium state of the Ice Giant magnetospheres?*** Voyager-2 spent only a few planetary rotations within Uranus' and Neptune's magnetopauses, such that it was challenging to establish a nominal configuration of their magnetospheres, their constituent particle populations, supporting current systems. Furthermore, it is unclear whether the observations of this dynamic system represent any kind of steady state, as ongoing magnetospheric reconfigurations were observed throughout each flyby, owing to the large dipole tilts at the Ice Giants and their ~16 and ~17-hour rotation periods. The extent to which the two magnetospheres are modified by internal plasma sources is also poorly constrained; Uranus' magnetosphere for instance was observed to be devoid of any appreciable cold plasma populations (McNutt et al., 1987). The presence of strong electron radiation belts (Mauk et al. 2010), seems contradictory to the absence of a dense, seed population at plasma energies, or could hint an efficient local acceleration process by intense wave-fields (Scarf et al. 1987). A strong proton radiation belt driven by Galactic Cosmic Ray (GCR)-planet collisions may reside closer to Uranus or Neptune than Voyager-2 reached (e.g. Stone et al. 1986), given that Earth and Saturn, which are known to sustain such belts, are exposed to a considerably lower GCR influx than the Ice Giants (Buchvarova and Belinov, 2009). Ion composition and UV aurora measurements hint that Triton could be a major source of plasma in Neptune's magnetosphere (Broadfoot et al., 1989, Richardson & McNutt, 1990), although questions remain as to the effects of coupling between the magnetosphere and the moon's atmosphere and ionosphere in establishing the plasma source (Hoogeveen & Cloutier, 1996). The magnetotails of both planets are expected to have very different structures to those seen at other magnetized planets (Figure 4), with strong helical magnetic field components (Cowley, 2013; Griton and Pantellini 2020), that may lead to a similarly helical topology of reconnection sites across the magnetospheric current sheet (Griton et al. 2018). Whether the overall magnetospheric configuration is controlled more by current sheet reconnection or a viscous interaction along the magnetopause (Masters et al. 2018) is also unknown. Finally, measuring average escape rates of ionospheric plasma would offer further insights on whether planetary magnetic fields protect planetary atmospheres from solar wind erosion (Wei et al. 2014, Gunell et al., 2018). For such dynamic systems, long-term observations that average out rotational effects and transients (e.g. Selesnick, 1988) are essential for achieving closure to all the aforementioned questions.

***How do the Ice Giant magnetospheres evolve dynamically?*** The large tilts of the Ice Giant magnetic fields relative to their planetary spin-axes hint that large-scale reconfigurations at diurnal time scales are dominating short-term dynamics, a view supported by MHD simulations (Griton and Pantellini 2020; Griton et al. 2018; Cao and Paty 2017; Mejnertsen et al. 2016). The rate of magnetic reconnection, for instance, is predicted to vary strongly with rotation, and so is the rate of matter and energy transfer into the magnetosphere and eventually upper atmosphere (Masters et al., 2014). Simulations do not capture how such variations impact regions as the radiation belts, which would typically require a stable environment to accumulate the observed, high fluxes of energetic particles (Mauk et al. 2010). A strong diurnal variability may also affect the space weather conditions at the orbits of the Ice Giant moons, regulating the interactions between the charged particle populations, their surfaces and exospheres (Plainaki et al., 2016; Arridge et al., 2014) through processes like surface sputtering, ion pick up, and charge exchange, which may also feed the magnetosphere with neutrals and low energy, heavy ion plasma (Lammer 1995). In the time-frame considered here, the exploration of Neptune could provide an opportunity to study a "pole-on" magnetosphere at certain rotational phases.

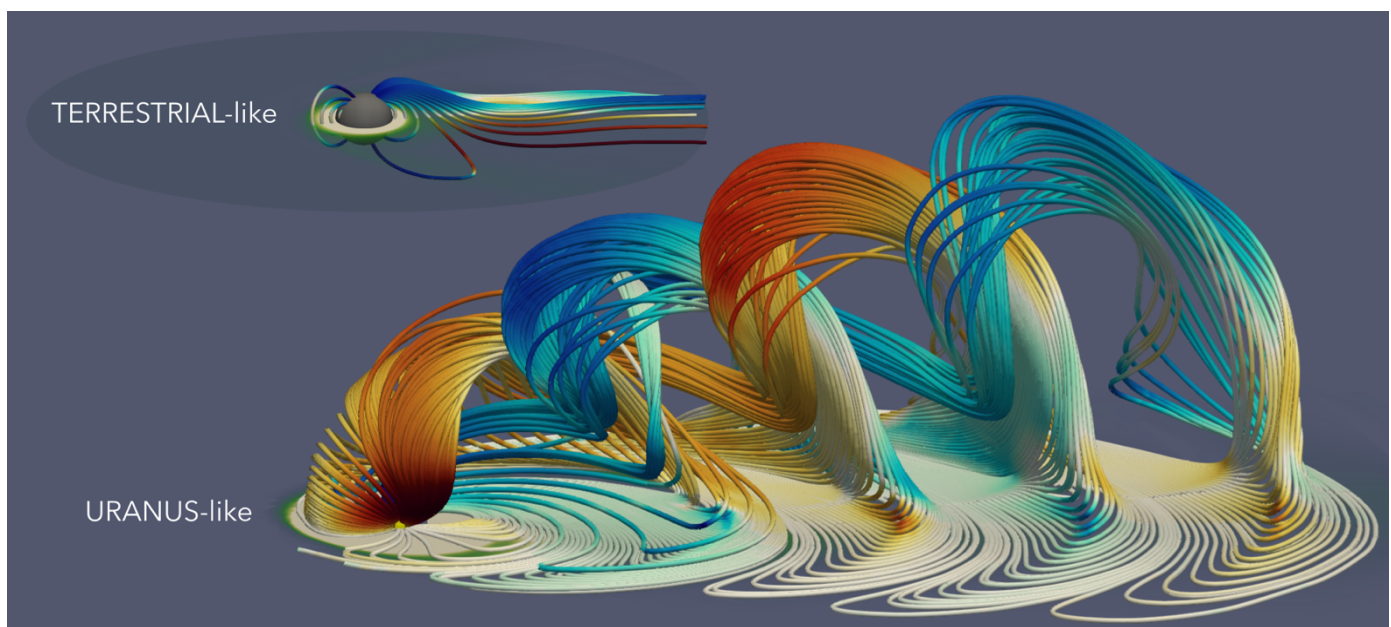

*Figure 4 Typical magnetic field configuration in a terrestrial-like, solar wind driven magnetosphere (top) where the magnetic and spin axis are almost aligned, compared to a Uranus-like magnetospheric configuration (bottom), when the magnetic axis is 90° away from the spin axis, during equinox. The helicoidal magnetotail structure develops due to the planet's fast rotation.*

Additional sources of variability can be solar wind driven, such as Corotating Interaction Regions (CIRs) and Interplanetary Coronal Mass Ejections (ICMEs) (Lamy et al. 2017). By the time they reach Uranus and Neptune, ICMEs expand and coalesce and attain a quasi-steady radial width of ~2-3 AU (Wang and Richardson, 2004) that could result in active magnetospheric episodes with week-long durations. With Triton being a potential active source, Neptune's magnetosphere may show variations at the moon's orbital period (Decker and Cheng 1994).

On longer time scales, seasonal changes of the planetary field's orientation are most important, especially at Uranus, because of its spin axis which almost lies on the ecliptic. There may be additional implications for the long-term variability of the magnetosphere if magnetic field measurements reveal a secular drift of the planetary field at any of the Ice Giants, in addition to providing constraints on the magnetic fields' origin and the planets' interiors (Section 2.1).

***How can we probe the Ice Giant magnetospheres through their aurorae?*** Aurorae form a unique diagnostic of magnetospheric processes by probing the spatial distribution of active magnetospheric regions and their dynamics at various timescales. Auroral emissions of Uranus and Neptune were detected by Voyager 2 at ultraviolet and radio wavelengths. The Uranian UV aurora has been occasionally re-detected by the Hubble Space Telescope since then. At NIR wavelengths auroral signatures remain elusive (e.g. Melin et al., 2019).

UV aurora are collisionally-excited H Ly-α and $H_2$ band emissions from the upper atmosphere, driven by precipitating energetic particles. The radical differences of the Uranian UV aurora observed across different seasons were assigned to seasonal variations of the magnetosphere/solar wind interaction (Broadfoot et al., 1986; Lamy et al., 2012, 2017; Cowley, 2013; Barthélémy et al., 2014). The tentative detection of Neptunian UV aurora did not reveal any clear planet-satellite interaction (e.g. with Triton) (Broadfoot et al., 1989). Repeated UV spectro-imaging observations are essential to probe the diversity of Ice Giant aurora, assess their underlying driver and constrain magnetic field models (e.g. Herbert, 2009).

Uranus and Neptune produce powerful auroral radio emissions above the ionosphere most likely driven by the Cyclotron Maser Instability as at Earth, Jupiter, and Saturn. The Uranian and Neptunian Kilometric Radiation are very similar (Desch et al., 1991, Zarka et al., 1995). They include (i) bursty emissions (lasting for <10min) reminiscent of those from other planetary magnetospheres. A yet-to-be-identified time-stationary source of free energy, able to operate in strongly variable magnetospheres, is thought to drive (ii) smoother emissions (lasting for hours) that are unique in our solar system (Farrell, 1992). Long-term remote and in-situ radio measurements are crucial to understand the generation of all types of Ice Giant radio emissions, to complete the baseline for the search of exoplanetary radio emissions. Long-term monitoring of auroral emissions will also be essential to precisely determine the rotation periods of these worlds.

**Summary*:* The Ice Giant magnetospheres comprise unique plasma physics laboratories, the study of which would allow us to observe and put to test a variety of astrophysical plasma processes that cannot be resolved under the conditions that prevail at the terrestrial planets and at the Gas Giants. The strong planet-magnetosphere links that exist further attest to the exploration of the magnetospheres as a key ingredient of the Ice Giant systems.**

## 2.4 Ice Giant Satellites – Natural & Captured

What can a comparison of Uranus' natural satellites, the captured "Ocean World" Triton, and the myriad ice-rich bodies in the Solar System, reveal about the drivers of active geology and potential habitability on icy satellites? The satellite systems of Uranus and Neptune offer very different insights into moon formation and evolution; they are microcosms of the larger formation and evolution of planetary systems. The Neptunian system is dominated by the 'cuckoo-like' arrival of Triton (i.e., severely disrupting any primordial Neptunian satellite system), which would be the largest known dwarf planet in the Kuiper belt if it were still only orbiting the Sun. Neptune's remaining satellites may not be primordial, given the degree of system disruption generated by Triton's arrival. In contrast, Uranus's satellite system appears to be relatively undisturbed since its formation — a highly surprising situation given that Uranus has the most severe axial tilt of any planet, implying a dramatic collisional event in its past. Thus, these two satellite systems offer laboratories for understanding the key planetary processes of formation, capture and collision.

***What can the geological diversity of the large icy satellites of Uranus reveal about the formation and continued evolution of primordial satellite systems?***
The five largest moons of Uranus (Miranda, Ariel, Umbriel, Titania, Oberon) are comparable in sizes and orbital configurations to the medium-sized moons of Saturn. However, they have higher mean densities, about 1.5 g/cm$^3$ on average, and have different insolation patterns: their poles are directed towards the Sun for decades at a time, due to the large axial tilt of Uranus. The surfaces of Uranus's five mid-sized moons exhibit extreme geologic diversity, demonstrating a complex and varied history (Figure 1). On Ariel and Miranda, signs of endogenic resurfacing associated with tectonic stress, and possible cryovolcanic processes, are apparent: these moons appear to have the youngest surfaces. Geological interpretation has suffered greatly from the incomplete Voyager 2 image coverage of only the southern hemisphere, and extremely limited coverage by Uranus-shine in part of the north (Stryk and Stooke, 2008). Apart from a very limited set of images with good resolution at Miranda, revealing fascinatingly complex tectonic history and possible re-formation of the moon (Figure 5), most images were acquired at low to medium resolution, only allowing characterisation of the main geological units (e.g., Croft and Soderblom, 1991) and strongly limiting any surface dating from the crater-size frequency distribution (e.g. Plescia, 1987a, Plescia, 1987b). High-resolution images of these moons, combined with spectral data, will reveal essential information on the tectonic and cryovolcanic processes and the relative ages of the different geological units, via crater statistics and sputtering processes. Comparison with Saturn's inner moons system will allow us to identify key drivers in the formation and evolution of compact multiplanetary systems.

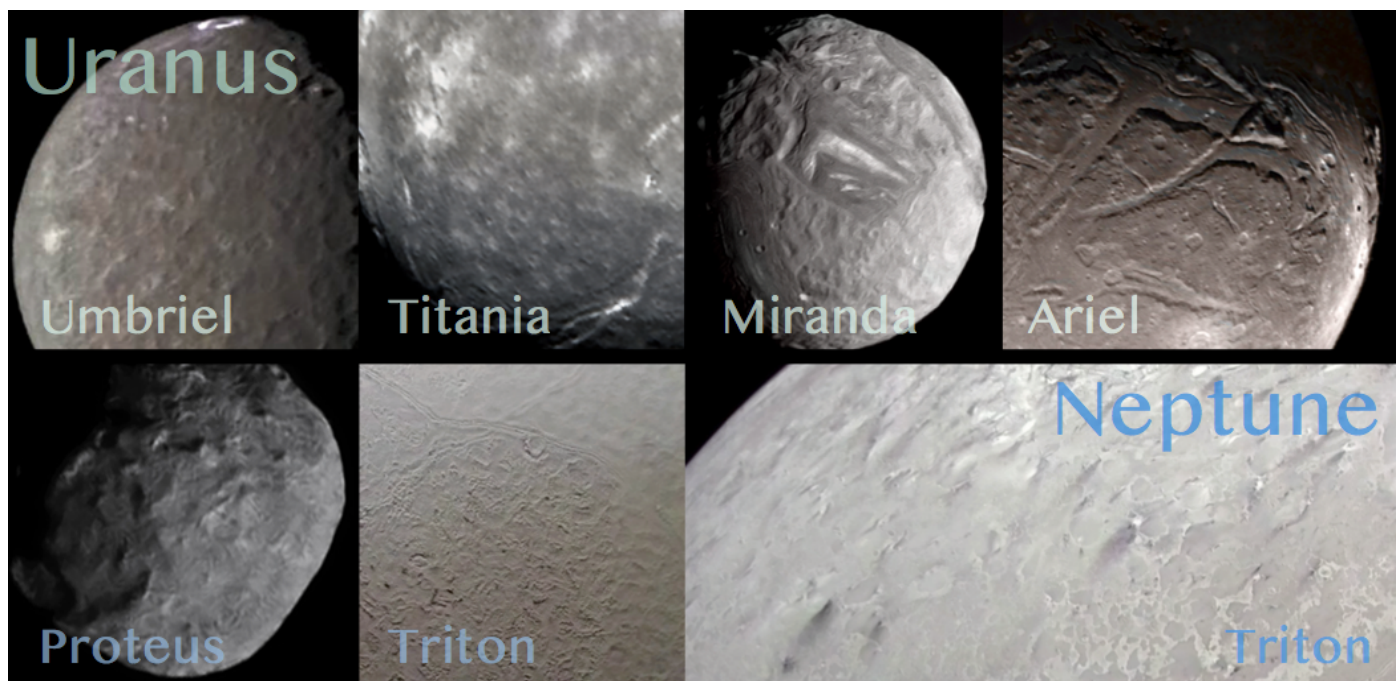

*Figure 5 Best-resolution (roughly ~1 km/px) imagery of terrains seen in the moons of Uranus (top row) and Neptune (lower row) by Voyager 2. At Uranus, Umbriel and Titania are highly cratered with some major faults; Miranda displays spectacular and massive tectonic features; Ariel's filled-fissured surface is suggestive of late cryovolcanic activity. At Neptune, Proteus has a surface suggestive of Saturn's dust-rich moon Helene. The dwarf-planet-sized Triton has both the sublimation-related "cantaloupe terrain" (left) and active nitrogen gas geysers in the south polar terrains (right), with deposited dust visible as dark streaks. Credit: NASA/JPL-Caltech/Ted Stryk/collage M. Bannister.*

***What was the influence of tidal interaction and internal melting on shaping the Uranian worlds, and could internal water oceans still exist?*** As in the Jovian and Saturnian systems, tidal interaction is likely to have played a key role in the evolution of the Uranian satellite system. Intense tidal heating during sporadic passages through resonances is expected to have induced internal melting in some of the icy moons (Tittemore and Wisdom 1990). Such tidally-induced melting events, comparable to those expected on Enceladus (e.g. Běhounková et al. 2012), may have triggered the geological activity that led to the late resurfacing of Ariel and possibly transient hydrothermal activity. The two largest (>1500 km diameter) moons, Titania and Oberon, may still harbour liquid water oceans between their outer ice shells and inner rocky cores – remnants of past melting events. Comparative study of the static and time-variable gravity field of the Uranian and Saturnian moons, once well-characterized, will constrain the likelihood and duration of internal melting events, essential to characterize their astrobiological potential. Complete spacecraft mapping of their surfaces could reveal recent endogenic activity.

Accurate radio tracking and astrometric measurements can also be used to quantify the influence of tidal interactions in the system at present, providing fundamental constraints on the dissipation factor of Uranus itself (Lainey, 2008). Gravity and magnetic measurements, combined with global shape data, will greatly improve the models of the satellites' interiors, providing fundamental constraints on their bulk composition (density) and evolution (mean moment of inertia). Understanding their ice-to-rock ratio and internal structure will enable us to understand if Uranus' natural satellite system are the original population of bodies that formed around the planet, or if they were disrupted.

***What is the chemical composition of the surfaces of the Uranian moons?*** The albedos of the major Uranian moons, considerably lower than those of Saturn's moons (except Phoebe and Iapetus' dark hemisphere), reveal that their surfaces are characterized by a mixture of $H_2O$ ice and a visually dark and spectrally bland material that is possibly carbonaceous in origin (Brown and Cruikshank, 1983). Pure $CO_2$ ice is concentrated on the trailing hemispheres of Ariel, Umbriel and Titania (Grundy et al., 2006), and it decreases in abundance with increasing semimajor axis (Grundy et al., 2006; Cartwright et al., 2015), as opposed to what is observed in the Saturnian system. At Uranus' distance from the Sun, $CO_2$ ice should be removed on timescales shorter than the age of the Solar System, so the detected $CO_2$ ice may be actively produced (Cartwright et al., 2015).

The pattern of spectrally red material on the major moons will reveal their interaction with dust from the decaying orbits of the irregular satellites. Spectrally red material has been detected primarily on the leading hemispheres of Titania and Oberon. $H_2O$ ice bands are stronger on the leading hemispheres of the classical satellites, and the leading/trailing asymmetry in $H_2O$ ice band strengths decreases with distance from Uranus. Spectral mapping of the distribution of red material and trends in $H_2O$ ice band strengths across the satellites and rings can map out infalling dust from Uranus's inward-migrating irregular satellites (Cartwright et al., 2018), similar to what is observed in the Saturnian system on Phoebe/Iapetus (e.g., Tosi et al., 2010), and could reveal how coupling with the Uranus plasma and dust environment influence their surface evolution.

***Does Triton currently harbour a subsurface ocean and is there evidence for recent, or ongoing, active exchange with its surface?*** Neptune's large moon Triton, one of a rare class of Solar System bodies with a substantial atmosphere and active geology, offers a unique opportunity to study a body comparable to the dwarf planets of the rest of the trans-Neptunian region, but much closer. Triton shares many similarities in surface and atmosphere with Pluto (Grundy et al. 2016), and both may harbour current oceans. Triton's retrograde orbit indicates it was captured, causing substantial early heating. Triton, in comparison with Enceladus and Europa, will let us understand the role of tidally-induced activity on the habitability of ice-covered oceans. The major discovery of plumes emanating from the southern polar cap of Triton (Soderblom et al. 1990; see Figure 5) by Voyager 2, the most distant activity in the Solar System, is yet to be fully understood (Smith et al. 1989).

Like Europa, Triton has a relatively young surface age of ~100 Ma (Stern and McKinnon 2000), inferred from its few visible impact craters. Triton also displays a variety of distinctive curvilinear ridges and associated troughs, comparable to those on Europa (Prockter et al. 2005) and especially apparent in the "cantaloupe terrain". This suggests that tidal stresses and dissipation have played an essential role in Triton's geological activity, and may be ongoing (Nimmo and Spencer 2015). Intense tidal heating following its capture could have easily melted its icy interior (McKinnon et al. 1995). Its young surface suggests that

Triton experienced an ocean crystallization stage relatively recently (Hussmann et al. 2006), associated with enhanced surface heat flux (Martin-Herrero et al. 2018). Combined magnetic, gravimetric and shape measurements from multiple flybys or in orbit will allow us to detect if Triton possesses an ocean and to constrain the present-day thickness of the ice shell. Correlating the derived shell structure and geological units will show if there is exchange with the ocean. It is entirely unclear if the source(s) for Triton's plumes reaches the ocean, as at Enceladus.

***Are seasonal changes in Triton's tenuous atmosphere linked to specific sources and sinks on the surface, including its remarkable plume activity?*** Triton's surface has a range of volatile ices seen in Earth-based near-infrared spectra, including $N_2$, $H_2O$, $CO_2$, and $CH_4$ (Quirico et al., 1999; Cruikshank et al., 2000; Tegler et al, 2012). A 2.239 μm absorption feature suggests that CO and $N_2$ molecules are intimately mixed in the ice rather than existing as separate regions of pure CO and pure $N_2$ deposits (Tegler et al., 2019). Mapping the spatial variation of this absorption feature will constrain how the surface-atmosphere interaction affects the surface composition, and more generally its climate and geologic evolution. Triton's surface may also contain complex organics from atmospheric photochemistry, like those of Pluto or Saturn's moon Titan (e.g. Krasnopolsky and Cruikshank 1995), as suggested by its yellowish areas (Thompson and Sagan 1990). Identifying the organic compounds, and mapping out their correlation with recently active terrains and geysers, will strongly raise the astrobiological potential of this exotic icy world.

Triton's tenuous atmosphere is mainly molecular nitrogen, with a trace of methane and CO near the surface (Broadfoot et al. 1989, Lellouch et al. 2010). Despite Triton's distance from the Sun and its cold temperatures, the weak sunlight is enough to drive strong seasonal changes on its surface and atmosphere. Because CO and $N_2$ are the most volatile species on Triton, they are expected to dominate seasonal volatile transport across its surface. Observation of increased $CH_4$ partial pressure between 1989 and 2010 (Lellouch et al. 2010) confirmed that Triton's atmosphere is seasonably variable. The plumes of nitrogen gas and dust could be a seasonal solar-driven process like the $CO_2$ `spiders' of the south polar regions of Mars, although an endogenic origin is possible.

***Are the smaller satellites of Neptune primordial?***
Voyager 2's flyby led to the discovery of six small moons inside Triton's orbit: Naiad, Thalassa, Despina, Galatea, Larissa and Proteus. A seventh inner moon, Hippocamp, has been recently discovered by HST observations orbiting between the two largest, Larissa and Proteus. Almost nothing is known about these faint moons, which may post-date Triton's capture rather than being primordial. Only a new space mission could unveil basic features such as shape and surface composition, shedding light on their origin.

***How does an Ice Giant satellite system interact with the planets' magnetospheres?*** Most of the major moons of Uranus and Neptune orbit within the planets' extensive magnetospheres (Figure 1). The tilt and offset of both planets' magnetic dipoles compared to their rotation axes mean that, unlike at Saturn, the major moons at both Ice Giants experience continually-changing external magnetic fields. The potential subsurface oceans of Titania, Oberon and Triton would be detectable by a spacecraft that can monitor for an induced magnetic field. The moons in both systems orbit in relatively benign radiation environments, but radiation belt particles could still drive sputtering processes at the inner moons' surfaces and Triton's atmosphere. Triton could be a significant potential source of a neutral gas torus and magnetospheric plasma at Neptune. Triton's ionosphere's transonic, sub-Alfvenic regime (Neubauer 1990; Strobel et al. 1990) may generate an auroral spot in Neptune's upper atmosphere. No such interaction is anticipated at Uranus. Red aurorae may also be present on Triton from $N_2$ emission, providing valuable insights into Triton's interaction with its space environment.

**Summary: Exploring Uranus' and Neptune's natural satellites, as well as Neptune's captured moon Triton, will reveal how ocean worlds may form and remain active, defining the extent of the habitable zone in the Solar System. Both satellite systems are equally compelling for future orbital exploration.**

## 2.5 Ice Giant Ring Systems

What processes shape the rings of Ice Giants, and why do they differ from the extensive rings of Saturn? Uranus and Neptune both possess a complex system of rings and satellites (Figure 6, see also de Pater et al., 2018, Nicholson et al., 2018). The rings exhibit narrow and dense ringlets, as well as fainter but broader dust components. The moons can confine the rings gravitationally, and may also serve as sources and sinks for ring material. Observations indicate rapid variability in the Uranian and Neptunian rings within decades. A mission to the Ice Giants, with a dedicated suite of instruments, can answer fundamental questions on the formation and evolution of the ring systems and the planets themselves:

***What is the origin of the solar system ring systems, and why are they so different?*** The origin of the giant planets' rings is one of the unsolved mysteries in planetary science. Whereas all four giant planets do have rings, their diversity of structure and composition argues for different formation scenarios (Charnoz et al., 2018). It was hypothesized that the very massive Saturnian rings formed more than 3 Gyrs ago through tidal destruction of a moon, or of a body on a path traversing the system. However, recent Cassini measurements (Zhang et al., 2017, Kempf et al., 2018, Waite et al., 2018, Iess et al., 2019) argue for a younger ring age. In contrast, Uranus' and Neptune's rings are far less massive and they have a different structure. Compared to Saturn, their rings' albedo is much lower, favouring a parent body that was a mixture of ice and dark material (silicates and possibly organics). For instance, it was suggested (Colwell and Esposito, 1992, 1993) that these two ring systems could result from the periodic destruction of moonlets though meteoroid bombardment, in which case most of the ringlets would be only transient structures, currently in the process of re-accretion to satellites. Among the Uranian dust rings the μ ring has a distinct blue spectral slope (de Pater et al. 2006). In that regard it is similar to Saturn's E ring, for which the blue slope results from the narrow size distribution of its grains, formed in the cryo-volcanic activity of the moon Enceladus. Although the moon Mab is embedded in the μ ring (Showalter et al, 2006) it appears much too small (12km radius) to be volcanically active and create in this way the dust that forms the ring. Other dust rings of Uranus exhibit a red spectral slope (de Pater et al. 2006), suggestive of dusty material released in micrometeoroid impacts on atmosphere-less moons and the origin for different appearance of the μ ring is unknown. Recent ground-based observations of thermal emission from the Uranian ring system (Molter et al., 2019) are consistent with the idea that the rings are made up of larger particles, without micron-sized dust, and that their temperatures result from low thermal inertia and/or slow rotation of the particles. Clearly, more data is needed to ultimately settle the question of the origin and nature of Uranus' and Neptune's rings.

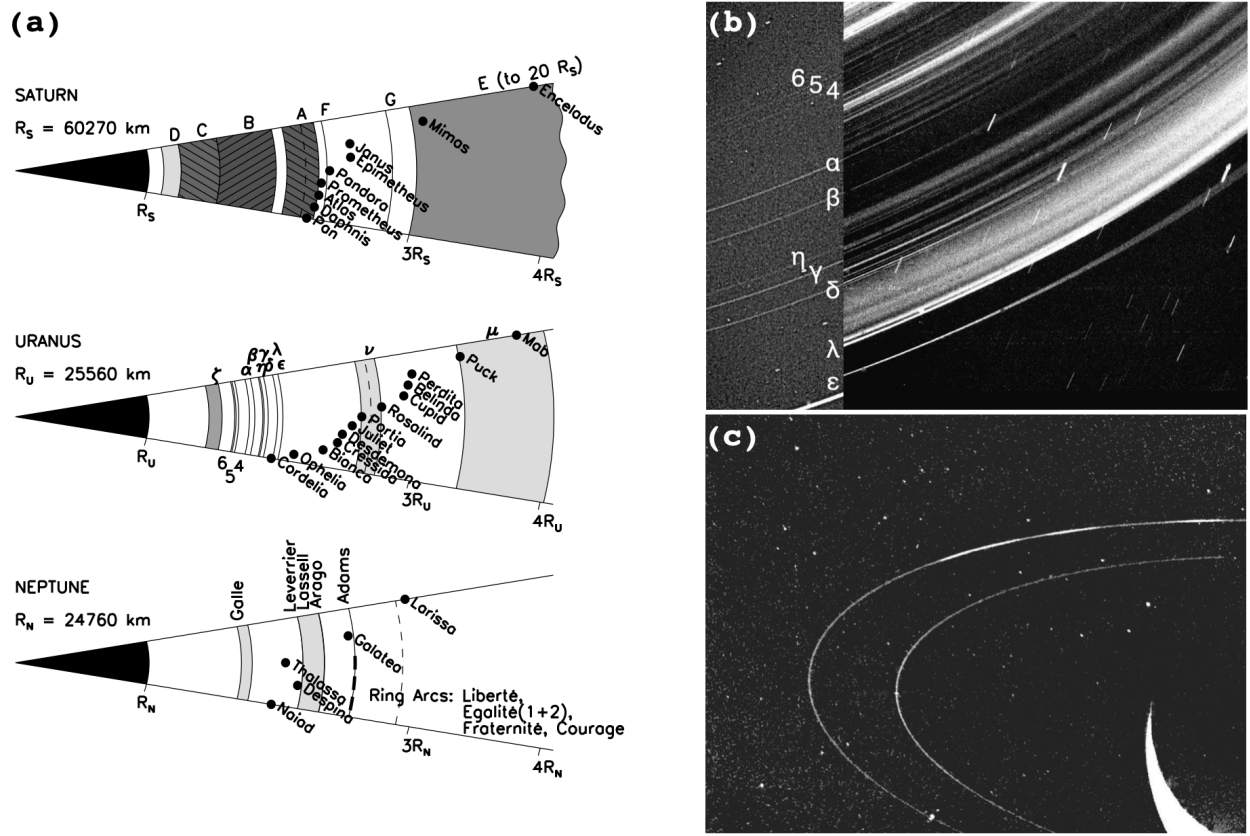

*Figure 6 Panel (a): Schematic diagram of the ring moon systems of Uranus, Neptune, and Saturn for comparison. The Roche radius (for icy ring particles) is marked as a dashed line. Panel (b) shows a combination of two Voyager images of the Uranus rings, taken at low (upper part) and high (lower part) phase angle (figure from Nicholson et al., 2018). At high phase angle the dusty components of the ring system stand out. Panel (c) shows a Voyager image of the rings of Neptune (Credit: NASA/JPL).*

***How do the ring-moon systems evolve?*** A variety of processes govern the evolution of planetary rings, many of which are also fundamental to other cosmic disks. Important for the rings are viscous transport, ring-satellite interactions, the self-gravity, as well as meteoroid bombardment and cometary impacts. These processes may induce rapid evolution on timescales much shorter than the rings' age. For instance, the Neptune ring arcs were initially interpreted to be confined by resonances with the moon Galatea. However, it was found that the arcs actually move away from the corresponding corotation sites (Dumas et al., 1999, Namouni & Porco 2002) and that they evolve rapidly (de Pater et al., 2005). Also, for the Uranian rings significant changes since the Voyager epoch are observed (de Pater et al., 2006, 2007). In the Uranus and Neptune systems the role the moons play in sculpting the rings is even stronger than for Saturn's rings. Moreover, depending on their composition, the Uranus and Neptune ring systems may even extend beyond the planet's Roche Limit (Tiscareno et al., 2013). This implies that their path of evolution is different from the Saturnian rings, inducing changes on more rapid timescales. Some of the edges of the Uranian rings are clearly confined by known satellites and others might be confined by yet undetected moons. Alternatively, there might be different processes of confinement at work, in a similar manner as for narrow rings of Saturn, for which shepherding moons are absent. Some of the dense rings of Uranus show sub-structure the origin of which is unclear. Spacecraft investigation will answer the question if the structure is induced by resonant interaction with moons, or if it represents intrinsic modes arising from instability and self-gravity of the rings.

***What is the ring composition?*** The rings (as the moons) very likely consist of material that was present at the location in the solar system where the planets themselves have formed. Therefore, the investigation of the composition of the rings, while being interesting in its own right, may also shed light on models of planet formation and migration in the solar system.

Imaging at high and intermediate phase angles will constrain the shapes and properties of known dust rings and has the potential to discover new rings. Multicolour imaging at a range of phase angles will determine the size distribution of the grains that form these rings. Imaging at low phase angles will probe the dense rings and allow for a comprehensive search and discovery of yet unseen satellites that serve as sources for ring material and that interact dynamically with the rings. Stellar occultations performed with a high-speed photometer, or radio occultations, will determine the precise optical depths of the denser rings and resolve their fine sub-structure (French et al., 1991). An IR spectrometer will determine the composition of the dense rings. In a complementary manner a dust detector will directly determine the composition of the grains forming the low optical depth dust rings (Postberg et al., 2009) and of particles lifted from the dense rings (Hsu et al., 2018). A dust detector will also measure the flux and composition of interplanetary particles in the outer solar system, which is a poorly known quantity of high importance for the origin and evolution of the rings.

**Summary: Ice Giant rings appear to be fundamentally different to those of Saturn, such that their origin, evolution, composition and gravitational relationships with the icy satellites should provide key new insights into the forces shaping ring systems surrounding planetary bodies.**

## 3. Ice Giant Science in Context

The scientific themes highlighted in Section 2 span multiple disciplines within planetary science, and address questions that touch on issues across astronomy. In this section we review how an Ice Giant mission must be considered in the context of other fields and technologies that will be developing in the coming decades.

### 3.1    Astronomical Observatories

An Ice Giant System mission would be operating in the context of world-leading new facilities in or near Earth. The 2020s will see the launch of the James Webb Space Telescope, able to provide spectral maps of both Ice Giants from 1-30 μm but at a moderate spatial resolution and with limited temporal coverage. Earth-based observatories in the 8-10 m class provide better spatial resolution at the expense of telluric obscuration. A successor to the Hubble Space Telescope, which has been the key provider of visible and UV observations of the Ice Giants, has yet to become a reality, but could be operating in the 2030s.    And although the next generation of Earth-based observatories in the 30+m class (the ELT, TMT, GMT) will provide exquisite spatial resolution, this will remain limited to atmospheric and ionospheric investigations of the Earth-facing hemisphere (with some disc-averaged spectroscopy of the satellites), leaving a multitude of fundamental questions unanswered.

Additionally, distant remote sensing can only study phenomena that alter the emergent spectrum – meteorology, seasonal variations, and ionospheric emissions (auroral and non-auroral).    This limits our understanding of the underlying mechanisms and means that ground- and space-based telescopes only serve a narrow subset of the Ice Giant community, and cannot address the wide-ranging goals described in Section 2.    Thus, there is no substitute for orbital exploration of one or both of these Ice Giant systems (alongside in situ sampling of their atmospheres), but we envisage that these space missions will work in synergy with the ground-based astronomy community, following successful examples of Galileo, Cassini, Juno and, ultimately, JUICE and Europa Clipper. Support from Earth-based observatories, either on the ground or in space, will be used to establish a temporal baseline for atmospheric changes (e.g., tracking storms), provide global context for close-in observations from the orbiters, and plug any gaps in spectral coverage or spectral resolution in the orbiter payload.

### 3.2    Heliophysics Connection

Missions to explore the Ice Giant Systems also resonate with the heliophysics community, as detailed exploration of an oblique rotator can inform a universal model of magnetospheres.    The panel on solar-wind magnetosphere interactions of the 2013 heliophysics decadal survey[5] identified how the magnetospheres of Uranus and Neptune are fundamentally different from others in our Solar System, and sought to ensure that magnetic field instruments would be guaranteed a place on outer planet missions, with a strong recommendation being that NASA's heliophysics and planetary divisions partner on a Uranus orbital mission. They describe how Uranus offers an example of solar wind/magnetosphere interactions under strongly changing orientations over diurnal timescales. Depending on the season, the effects of solar wind-magnetosphere interaction vary dramatically over the course of each day.

There is also a need to understand how the solar wind evolves beyond 10 AU (Saturn orbit), as the states of solar structures travelling within the solar wind (solar wind pressure pulses) are largely unknown due to the lack of observations at such large heliocentric distances (Witasse et al., 2017). The long cruise duration of a mission to Uranus or Neptune provides an excellent opportunity for both heliophysics and Ice Giant communities if a space weather-monitoring package that includes a magnetometer, a solar wind analyser, and a radiation monitor operates continuously during the cruise phase, to understand how conditions vary out to 20-30 AU over a prolonged lifetime. Moreover, the complexity of the interactions of Uranus's magnetosphere with the solar wind provides an ideal testbed for the theoretical understanding of planetary interactions with the solar wind, significantly expanding the parameter range over which scientists can study magnetospheric structure and dynamics. The potential discoveries from its dynamo generation and its variability stand to open new chapters in comparative planetary magnetospheres and interiors.

### 3.3 Exoplanet & Brown Dwarf Connection

The Ice Giant System mission will occur during an explosion in our understanding of planets beyond our Solar System, through ESA's Cosmic Vision missions

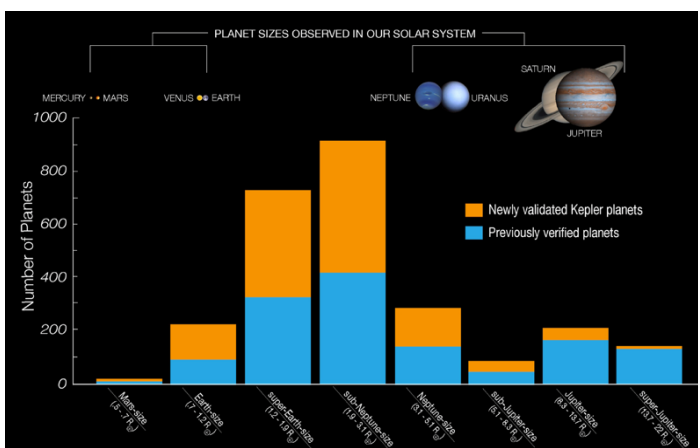

*Figure 7 Known transiting exoplanets in 2016, from the Kepler mission, showing sub-Neptunes as the most common planetary radius in the current census. Credit: NASA/Ames*

Plato, Euclid, and ARIEL; through missions with international partners like JWST and TESS; and through next-generation observatories like WFIRST, LUVOIR, Origins, and HabEx. The physical and chemical processes at work within our own Solar System serve as the key foundation for our understanding of those distant and unresolved worlds. Our Solar System provides the only laboratory in which we can perform in-situ experiments to understand exoplanet formation, composition, structure, dynamos, systems and magnetospheres. After the several highly-successful missions to the Gas Giants (and the upcoming ESA/JUICE mission), dedicated exploration of the Ice Giants would place those discoveries into a broader, solar-system context.

***Planet Statistics:*** Uranus and Neptune represent our closest and best examples of a class of planets intermediate in mass and size between the larger, hydrogen-helium-enriched gas giants, and the rocky terrestrial worlds. Indeed, Neptune- and sub-Neptune-size worlds have emerged as commonplace in our ever-expanding census of exoplanets (Figure 7). Neptune-size planets are among the most common classes of exoplanet in our galaxy (Fulton et al., 2018). Fressin et al. (2013) suggest that this category of planets can be found around 3-31% of sun-like stars, and Petigura et al (2017) suggests that sub-Neptunes remains the most common category within Kepler's survey. Based on statistics from the Kepler mission it is predicted that TESS will detect over 1500 sub-Neptunes (2-4 $R_E$) over the mission (Barclay et al. 2018). Microlensing surveys (e.g., WFIRST, Penny et al., 2019) will also be more sensitive to lower-mass planets on wide orbits, and could reveal new insights into extrasolar Ice Giants ahead of a mission to Uranus or Neptune. Given these planetary occurrence rates, the exploration of bulk composition and interiors of our Ice Giants would provide strong constraints on the most common outcome for planetary formation. However, we emphasise that being Neptune-sized does not necessarily imply being Neptune-like, as a plethora of additional parameters come into play to shape the environmental conditions on these worlds.

***Atmospheric Insights:*** Although we are currently unable to resolve spatial contrasts on exoplanets and brown dwarfs, comparisons of dayside (eclipse) and nightside (transit) spectra suggest the presence of powerful winds on some targets that are responsible for redistribution of energy. Brightness variations as Brown Dwarfs rotate suggest patchy cloud structures, but their rapid evolution was only understood when compared with long-cadence Neptune observations (Apai et al., 2013, Stauffer et al. 2016, Simon et al. 2016). Changes in exoplanet transit spectra with temperature indicate complex cloud condensation processes (Morley et al., 2012; Wakeford et al., 2017). Atmospheric dynamics, chemistry, and cloud formation all vary as a function of planetary age, distance from the host star, and the bulk enrichments of chemical species. The smaller radii of Uranus and Neptune, compared to the larger gas giants, implies atmospheric processes (zonal banding, storms, vortices) at work in a different region of dynamical parameter space, and one which is unavailable elsewhere in our Solar System. Detailed exploration of our Ice Giants, in comparison to the existing studies of the Gas Giants, would allow us to unravel these competing and complex phenomena shaping the atmospheres of exoplanets and brown dwarfs. Importantly, measurements of Ice Giant composition and dynamics can be directly compared to exoplanet and brown dwarf observations, placing their formation location, age, and evolution into context, and vice versa. However, this cannot be done without a detailed comparative dataset from our Solar System Ice Giants.

**Magnetospheric Insights:** Rymer et al. (2018) explore the importance of Ice Giant interactions with the wider magnetic environment as a testbed for understanding exoplanets. Eccentric and complex orbital characteristics appear commonplace beyond our Solar System, and Uranus is one of the only places where radio emission, magnetospheric transport and diffusion resulting from a complex magnetospheric orientation can be explored. The stability and strength of the Uranian radiation belts could also guide the search for exoplanets with radiation belts. Finally, understanding the dynamos of Uranus and Neptune would drastically improve our predictions of magnetic field strengths and exoplanet dynamo morphologies. Each of these insights will be vital as exoplanetary science moves into an era of characterisation of atmospheric composition, dynamics, clouds, and auroral/radio emissions.

# 4. Ice Giant Missions

## 4.1 Architectures: The Case for Orbiters

The scientific themes of Ice Giant System missions (summarised in Figure 9) are broad and challenging to capture within a single mission architecture, but recent efforts by both ESA (e.g., the 2018-19 studies with the

Concurrent Design Facility for an M*-class mission[6]) and NASA (e.g., the 2017 Science Study Team report, Hofstadter et al., 2019)[7] have explored strategies to achieve many of the goals in Section 2. The joint NASA-ESA science study team provided a detailed investigation of various combinations of flyby missions, orbiters, multiple sub-satellites from a core spacecraft, satellite landers, and atmospheric probes. Strategies to explore both Ice Giants with dual spacecraft have also been proposed (Turrini et al., 2014; Simon et al., 2018). It was widely recognised that a flyby mission like Voyager, without any additional components like an entry probe, was deemed to provide the lowest science return for the Ice Giants themselves, despite their lower cost point. Without *in situ* measurements, and by providing only brief snapshots of the evolving atmospheres and magnetospheres, and limited coverage of the satellites and rings, a flyby could not deliver on the highest-priority science goals for an Ice Giant mission. Targeting Triton as a flyby, or the inclusion of an entry probe, would increase the scientific reach, but would still prove inadequate for whole-system science. The study found that an Ice Giant orbiter for either the Uranus or Neptune systems, alongside an *in situ* atmospheric probe, would provide an unprecedented leap in our understanding of these enigmatic worlds. An orbiter would maximise the time spent in the system to conduct science of interest to the entire planetary community.

In our 2019 submission to ESA's call for ideas to shape the planning of space science missions in the coming decades (known as Voyage 2050), we therefore proposed that an orbital mission to an Ice Giant should be considered as a cornerstone of ESA's Voyage 2050 programme, if not already initiated with our international partners in the coming decade. An ESA orbital mission, powered by radioisotope thermoelectric generators, should be studied as an L-class mission to capitalise on the wealth of European experience of the Cassini and JUICE missions. Alternatively, an M-class Ice Giant budget would allow a crucial contribution to an orbital mission led by our international partners. The mass and mission requirements associated with additional components, such as satellite landers, *in situ* probes, or secondary small satellites, must be tensioned against the capabilities of the core payload, and the capability of the launch vehicle and propulsion. In all of these cases, a formal study of the requirements and capabilities is necessary to mature the concept.

**Payload Considerations:** The 2017 NASA study found that payload masses of 90-150 kg could deliver significant scientific return for a flagship-class mission, whilst the 2018 ESA CDF study identified 100 kg as a realistic payload mass for a European orbiter. Different studies have resulted in different prioritisations for instrumentation, but produced suites of orbiter experiments in common categories. Multi-spectral remote sensing is required, using both imaging and spectroscopy, spanning the UV, visible, near-IR (e.g., atmosphere/surface reflectivity, dynamics; auroral observations), mid-IR, sub-mm, to centimetre wavelengths (e.g., thermal emission and energy balance, atmospheric circulation). Such remote sensing is also a requirement for characterising any atmospheric probe entry sites, or satellite lander sites. Direct sensing of the magnetospheric and plasma environment would be accomplished via magnetometers, dust detectors, plasma instruments, radio wave detectors, and potentially mass spectrometers. Radio science would provide opportunities for interior sounding and neutral/ionospheric occultation studies. The provision of such instruments would capitalise on European heritage on Cassini, JUICE, Rosetta, Venus/Mars Express, and BepiColombo, but at the same time recognising the need to develop smaller, lighter, and less power/data-intensive instruments, raising and maturing their technological readiness.

**Orbit Considerations:** Orbital missions to both Uranus and Neptune depend crucially on the chemical fuel required for orbit insertion, which determines the deliverable mass. The potential use of aerocapture, using atmospheric drag to slow down the spacecraft, permits larger payloads and faster trip times at the expense of increased risk, which needs significant further study. Mission requirements and orbital geometries will determine the inclination of orbital insertion – high geographical latitudes would benefit some atmospheric, rings, and magnetospheric science, but satellite gravity assists and subsequent trajectory corrections would be needed for exploration of the satellites, rings and atmosphere from a low-inclination orbit. High inclinations are easier to achieve at Uranus, although Triton can be used to drive a satellite tour at Neptune. The delivery and telemetry for an atmospheric entry probe must also be considered in an orbital tour design (e.g., Simon et al., 2020). We also propose that distinct phases of an orbital tour be considered, balancing moderate orbital distances (for remote sensing, outer magnetosphere science, and a satellite tour) with close-in final orbits (for gravity science and inner magnetosphere), following the example of Cassini and Juno. Multiple close flybys of major satellites are desirable to map their interiors, surfaces, and atmospheres via a variety of techniques. Finally, multi-year orbital tours (at least ~3 years) would maximise our time in the system, permitting the study of atmospheric and magnetospheric changes over longer time periods. The 2018 ESA CDF study confirmed the feasibility of orbital tours satisfying these scientific requirements.

**Ring Hazards:** The 2017 NASA-ESA report highlighted potentially unknown ring-plane hazards as a topic for future study. Orbit insertion should be as close to the planet as possible to reduce the required fuel, but the properties of Ice Giant rings remain poorly constrained. Potential options to mitigate this risk include: having the insertion be further out (requiring more fuel); fly through the ring plane at an altitude where atmospheric drag is high enough to reduce the number of particles, but not enough to adversely affect the spacecraft; use a pathfinder spacecraft to measure the particle density ahead of time; or use Earth-based observations to constrain the upper atmosphere/ring hazard. Detailed calculations on the location of this safe zone are required.

## 4.2    Timeliness and Launch Opportunities

Trajectories to reach the Ice Giants depend on a number of factors: the use of chemical and/or solar-electric propulsion (SEP) technologies; the lift capacity of the launch vehicle; the use of aerocapture/aerobraking; and the need for gravity assists. The availability of Jupiter, as the largest planet, is key to optimal launch trajectories, and the early 2030s offer the best opportunities. The synodic periods of Uranus and Neptune with respect to Jupiter are ~13.8 and ~12.8 years, respectively, meaning that optimal Jupiter gravity-assist (GA) windows occur every 13-14 years. The NASA-ESA joint study team identified chemical-propulsion opportunities with a Jupiter GA in 2029-30 for Neptune, and a wider window of 2030-34 for Uranus. Such windows would repeat in the 2040s, and a wider trade space (including the potential to use Saturn GA or direct trajectories using the next generation of launch vehicles) should be explored. We stress that a mission to an Ice Giant is feasible using conventional chemical propulsion.

Furthermore, a mission launched to Uranus or Neptune could offer significant opportunities for flybys of Solar System objects en route, especially Centaurs, which are small bodies that orbit in the giant planet region. This population has yet to be explored by spacecraft, but represents an important evolutionary step between Kuiper Belt objects and comets. Around 300 are currently known, with 10% of these observed to show cometary activity. In addition, we can expect to discover at least an order of magnitude more Centaurs by the 2030s, following discoveries by the Large Synoptic Survey Telescope, increasing the probability that a suitable flyby target can be found near to the trajectory to Uranus/Neptune. Some of the largest of the Centaurs (~100 km scale objects) have their own ring systems, the origin of which has yet to be explained, while smaller ones (1-10 km scale) could add an important 'pre-activity' view to better interpret data from Rosetta's exploration of a comet. The payload options described in Section 4 would be well suited to characterise a Centaur during a flyby.

The launch time necessarily influences the arrival time, as depicted in Figure 8. Uranus will reach northern summer solstice in 2030, and northern autumnal equinox in 2050. Voyager 2 observed near northern winter solstice, meaning that the north poles of the planet and satellites were shrouded in darkness. These completely unexplored northern terrains will begin to disappear into darkness again in 2050, where they will remain hidden for the following 42 years (half a Uranian year).

Neptune passed northern winter solstice in 2005, and will reach northern spring equinox in 2046. After this time, the southern hemispheres of the planet and satellites (most notably the plumes of Triton at high southern latitudes) will sink into winter darkness, meaning that the Triton plumes – if they are indeed restricted to the south – would no longer be in sunlight after ~2046, and would remain hidden for the next ~82 years (half a Neptunian year).

The 2028-34 launch opportunities were assessed by the joint ESA-NASA study team. Saturn GA was considered but did not appear optimal for this launch window. Interplanetary flight times are 6 to 12 years to Uranus, 8 to 13 years to Neptune, depending on launch year, mission architecture, and launch vehicle. The greater challenge of reaching the Neptune system was reflected in their choice of detailed architecture studies: five missions to Uranus (orbiters with/without probes; with/without SEP; and with different payload masses), and a single orbiter and probe for Neptune. Both Uranus and Neptune were deemed equally valuable as targets – Uranus standing out for its uniqueness; Neptune for the prospects of exploring Triton. The choice between these two enticing destinations will ultimately be driven by launch opportunities and deliverable mass to the systems.

We must capitalise on the current momentum within Europe, alongside our international partners, to make use of the launch opportunities in the ~2030s. Such a mission would arrive at Uranus while we can still see the totally-unexplored northern terrains, or at Neptune while we can still see the active geology of Triton. Operations in the 2040s would also allow ESA to maintain Outer Solar System expertise from current and future missions like JUICE. An Ice Giant explorer would therefore be active as a cornerstone of ESA's Voyage 2050 strategic planning for space missions.

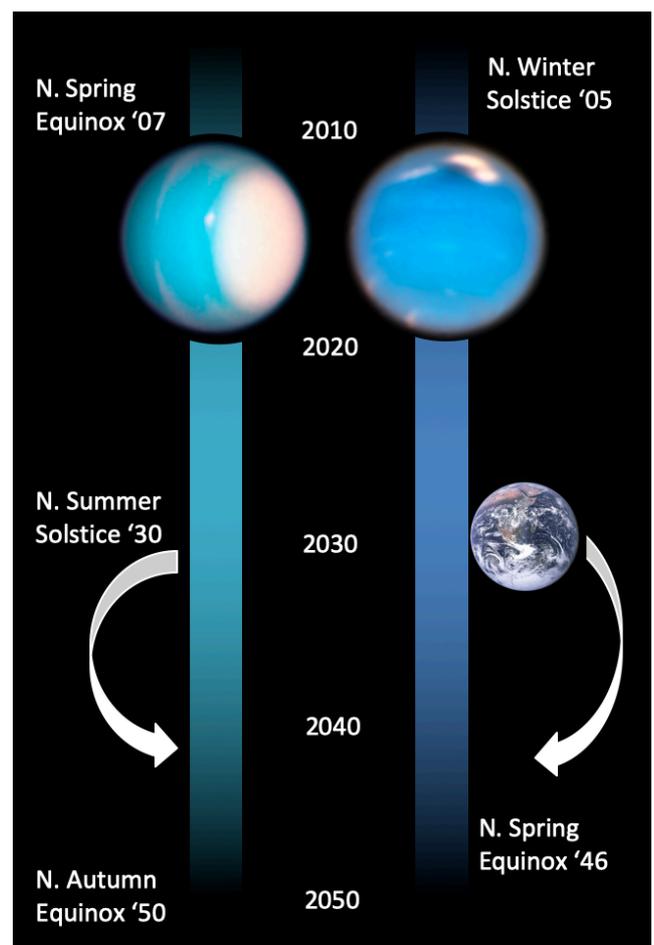

*Figure 8 Potential timeline of missions to Uranus (left) and Neptune (right), compared to the seasons on each Ice Giant. The white arrows show the approximate timescales for launch opportunities in the early 2030s, with arrival in the 2040s.*

### 4.3    Mature and Developing Technologies

An ambitious mission to an Ice Giant System would largely build on existing mature technologies (e.g., see the discussion of payload development in Section 4.1), but several challenges have been identified that, if overcome, would optimise and enhance our first dedicated orbital mission to these worlds. Note that we omit the need for ablative materials on atmospheric entry probes, which will be required for *in situ* science (Mousis et al., 2018, Simon et al., 2020). Key areas where technology maturations are required include:

**Space nuclear power:** With the prospect of flying solar-powered spacecraft to 20 AU being non-viable, an Ice Giant System mission must rely on radioisotope power sources, both for electricity and for spacecraft heating. In the US, existing MMRTGs (multi-mission radioisotope thermal generators), based on the decay of $^{238}$Pu, will be re-designed to create eMMRTGs ("enhanced") to increase the available specific power at the end of life, 4-5 of which were considered for the mission architectures studied in the 2017 NASA-ESA report. Previous M-class Uranus mission proposals have relied on US provision of these power sources for an ESA-led mission. However, ESA continues to pursue the development of independent power sources based on $^{241}$Am (Ambrosi et al., 2019). Whilst the power density is lower than that of $^{238}$Pu, the half-life is much longer, and much of the material is available from the reprocessing of spent fuel from European nuclear reactors, extracted chemically from plutonium to a ceramic oxide form. Prototypes for both radioisotope heater units (warming the spacecraft) and thermoelectric generators (providing spacecraft power) have now been demonstrated, and development is continuing for operational use late in the next decade (Ambrosi et al., 2019). An Ice Giant System mission could benefit tremendously from this independent European power source.

**Hardware longevity:** Given the 6- to 13-year interplanetary transfer, coupled with the desire for a long orbital tour, Ice Giant orbiters must be designed to last for a long duration under a variable thermal load imposed by gravity assists from the inner to the outer solar system. This poses constraints on the reliability of parts and power sources, as well as the need to develop optimised operational plans for the long cruise phases, such as the use of hibernation modes following the example of New Horizons.

**Telemetry/Communications:** All missions to the giant planets are somewhat constrained by competition between advanced instrumentation and the reduced data rates at large distances from Earth, but the case at Uranus and Neptune is most severe. The use of Ka-band in the downlink, currently supported by both the US Deep Space Network and two out of three ESA Deep Space Antennas, allows for mitigating this issue by increasing the achievable daily data volumes, but a careful optimization of the science tour remains critical. The available power, length of the downlink window, and the need to balance science data, engineering and housekeeping telemetry, all contribute to determining the overall data rate. For example, the NASA/JPL mission study[8] in 2017 suggested that use of a 35-W traveling wave tube amplifier (the current power limit for space-qualified TWTAs) and a 34-m ground-station could provide 15 kbps at Uranus, whereas this could increase to 30 kbps with a 70-W TWTA. The 2018 ESA CDF study[9] suggested that the data rate of a future M-class orbital mission to Uranus could use a 100-W TWTA to deliver 94 kbps Ka-band downlink from Uranus, or 42 kbps from Neptune. Assuming 3.2 hours/day for communications, this equates to daily data volumes for science of 1.09 and 0.48 Gb, respectively. Using current 35-W TWTAs, the ESA CDF study also suggested that lower data rates of 31 and 14 kbps could be achievable at Uranus and Neptune respectively, requiring the use of longer downlink windows. Even under the most optimistic assumptions discussed above, it is clear that some data optimization strategy would still be required. We would welcome detailed studies of new communications technologies, such as optical communications, as a general enabling technology for solar system exploration. However, we recognise that achieving the required directionality of a downlink laser from beyond 5AU will be challenging.

**Launch Vehicles:** The market for launch vehicles is changing dramatically both in Europe and in the US. Ice Giant mission concepts have traditionally considered Jupiter gravity assists to provide realistic flight times and sufficient payload delivered to each system.

---

[8] https://www.lpi.usra.edu/icegiants/mission_study/Full-Report.pdf

[9] https://sci.esa.int/web/future-missions-department/-/61307-cdf-study-report-ice-giants

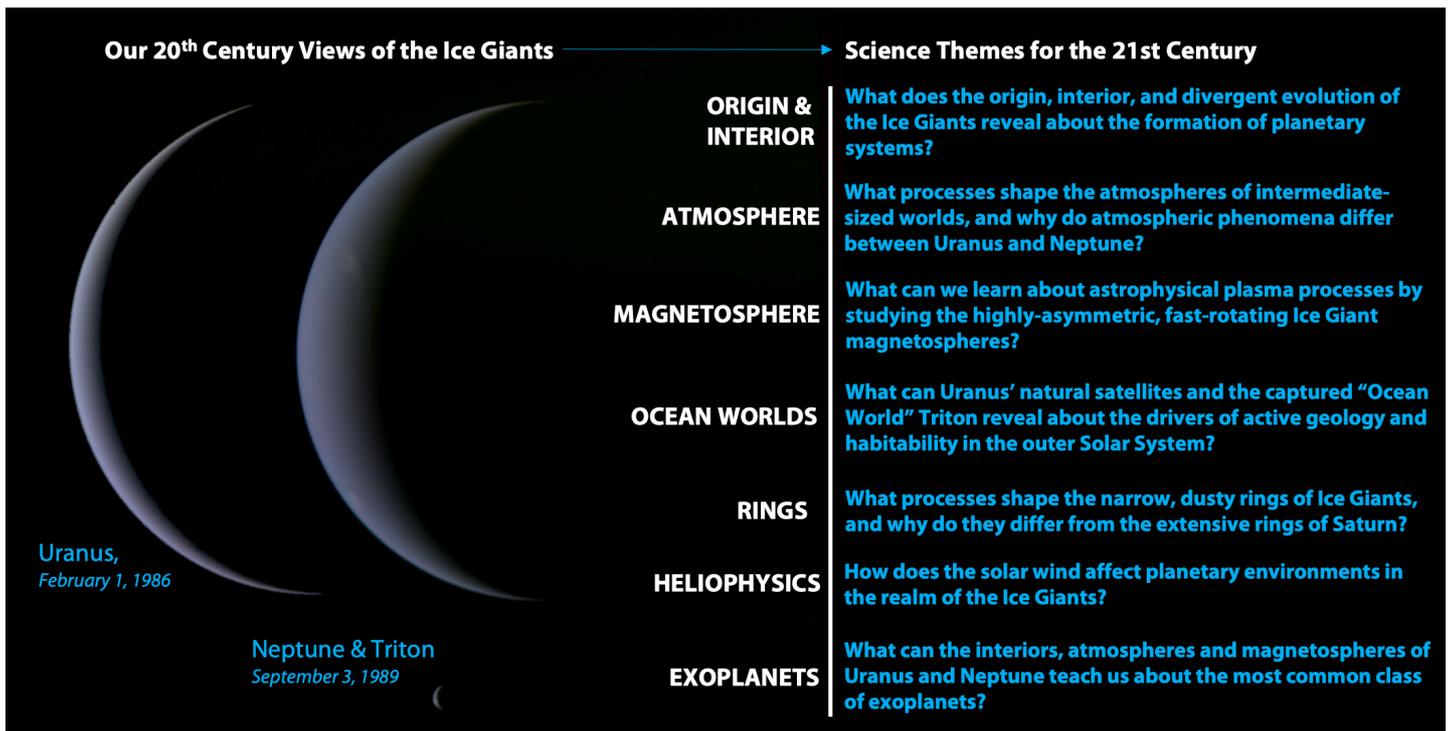

*Figure 9 Left: Our last views of Uranus and Neptune from a robotic spacecraft, taken by Voyager 2 three decades ago (Credit: NASA/JPL/E. Lakdawalla).  Will we see these views again before a half-century has elapsed?  Right:  Seven science themes for Voyage 2050 that could be addressed by an Ice Giant System mission.*

However, we also advocate investigation of the direct-transfer trajectories enabled by the heavy-lift capacity of the next generation of launch vehicles.  This may (i) obviate the need for Jupiter/Saturn GA, so allow more flexibility in launch dates; and (ii) open up the possibilities for launching multiple spacecraft that share the same faring.

## 5. Summary and Perspectives

This article reviews and updates the scientific rationale for a mission to an Ice Giant system, advocating that an orbital mission (alongside an atmospheric entry probe, Mousis et al., 2018) be considered as a cornerstone of ESA's Voyage 2050 programme, working in collaboration with international partners to launch the **first dedicated mission** to either Uranus or Neptune.  Using technologies both mature and in development, the Ice Giants community hopes to capitalise on launch opportunities in the 2030s to reach the Ice Giants.  As shown in Figure 9, an Ice Giant System mission would engage a wide community, drawing expertise from a vast range of disciplines *within* planetary science, from surface geology to planetary interiors; from meteorology to ionospheric physics; from plasma scientists to heliophysicists.  But this challenge is also interdisciplinary in nature, engaging those studying potentially similar Neptune-size objects beyond our

Solar System, by revealing the properties of this underexplored class of planetary objects.  As Neptune's orbit shapes the dynamical properties of objects in the distant solar system, an Ice Giant System mission also draws in the small-bodies community investigating objects throughout the Outer Solar System, from Centaurs, to TNOs, to Kuiper Belt Objects, and contrasting these with the natural satellites of Uranus.

To launch a mission to Uranus and/or Neptune in the early 2030s, the international Ice Giant community needs to significantly raise the maturity of the mission concepts and instrument technological readiness.  At the time of writing (late 2019), we await the outcomes of both ESA's Voyage 2050 process, and the next US planetary decadal survey.  Nevertheless, this should not delay the start of a formal science definition and study process (pre-phase A conceptual studies), so that we are ready to capitalise on any opportunities that these strategic planning surveys provide.  Fully costed and technologically robust mission concepts need to be developed, studied, and ready for implementation by 2023-25 to have the potential to meet the upcoming window for Jupiter gravity assists between 2029-2034.  We hope that the scientific themes and rationale identified in this review will help to guide that conceptual study process, to develop a paradigm-shifting mission that will help redefine planetary science for a generation of scientists and engineers.

Most importantly, the Ice Giant System mission will continue the breath-taking legacy of discovery of the Voyager, Galileo, Cassini, and Juno missions to the giant planets (in addition to the forthcoming JUICE and Europa Clipper missions). A dedicated orbiter of an Ice Giant is the next logical step in our exploration of the Solar System, completing humankind's first reconnaissance of the eight planets. It will be those discoveries that no one expected, those mysteries that we did not anticipate, and those views that no human has previously witnessed, which will enthuse the general public, and inspire the next generation of explorers to look to the worlds of our Solar System. We urge both ESA and NASA to take up this challenge.

## Acknowledgements

Fletcher was supported by a Royal Society Research Fellowship and European Research Council Consolidator Grant (under the European Union's Horizon 2020 research and innovation programme, grant agreement No 723890) at the University of Leicester. Mousis acknowledges support from CNES. Hueso was supported by the Spanish MINECO project AYA2015-65041-P (MINECO/FEDER, UE) and by Grupos Gobierno Vasco IT1366-19 from Gobierno Vasco. UK authors acknowledge support from the Science and Technology Facilities Council (STFC). The authors are grateful to Beatriz Sanchez-Cano, Colin Snodgrass, Hannah Wakeford, Marius Millot, and Radek Poleski for their comments on early drafts of this article, and to two anonymous reviewers for their helpful suggestions. We are grateful to the members of ESA's Voyage 2050 Senior Committee for allowing us to present this review at the Voyage 2050 workshop in Madrid in October 2019 (see https://www.cosmos.esa.int/web/voyage-2050/workshop-programme).

## References

Ambrosi, R. M., Williams, H., Watkinson, E. J., et al. (2019), European Radioisotope Thermoelectric Generators (RTGs) and Radioisotope Heater Units (RHUs) for Space Science and Exploration, Space Science Reviews, 215, 55, 10.1007/s11214-019-0623-9.

Apai, D., Radigan, J., Buenzli, E., et al. (2013), HST Spectral Mapping of L/T Transition Brown Dwarfs Reveals Cloud Thickness Variations, The Astrophysical Journal, 768, 121, 10.1088/0004-637X/768/2/121.

Arridge, C. S., Achilleos, N., Agarwal, J., et al. (2014), The science case for an orbital mission to Uranus: Exploring the origins and evolution of ice giant planets, Planetary and Space Science, 104, 122, 10.1016/j.pss.2014.08.009.

Arridge, C. S., Agnor, C. B., André, N., et al. (2012), Uranus Pathfinder: exploring the origins and evolution of Ice Giant planets, Experimental Astronomy, 33, 753, 10.1007/s10686-011-9251-4.

Atreya, S. K., Hofstadter, M. H., In, J. H., et al. (2020), Deep Atmosphere Composition, Structure, Origin, and Exploration, with Particular Focus on Critical in situ Science at the Icy Giants, Space Science Reviews, 216, 18, 10.1007/s11214-020-0640-8.

Bailey, E., & Stevenson, D. J. (2015), Modeling Ice Giant Interiors Using Constraints on the $H_2$-$H_2O$ Critical Curve, AGU Fall Meeting Abstracts, 2015, P31G-03.

Bali, E., Audétat, A., & Keppler, H. (2013), Water and hydrogen are immiscible in Earth's mantle, Nature, 495, 220, 10.1038/nature11908.

Barclay, T., Pepper, J., & Quintana, E. V. (2018), A Revised Exoplanet Yield from the Transiting Exoplanet Survey Satellite (TESS), The Astrophysical Journal Supplement Series, 239, 2, 10.3847/1538-4365/aae3e9.

Barthélemy, M., Lamy, L., Menager, H., et al. (2014), Dayglow and auroral emissions of Uranus in $H_2$ FUV bands, Icarus, 239, 160, 10.1016/j.icarus.2014.05.035.

Batalha, N. M., Rowe, J. F., Bryson, S. T., et al. (2013), Planetary Candidates Observed by Kepler. III. Analysis of the First 16 Months of Data, The Astrophysical Journal Supplement Series, 204, 24, 10.1088/0067-0049/204/2/24.

Běhounková, M., Tobie, G., Choblet, G., et al. (2012), Tidally-induced melting events as the origin of south-pole activity on Enceladus, Icarus, 219, 655, 10.1016/j.icarus.2012.03.024.

Boué, G., & Laskar, J. (2010), A Collisionless Scenario for Uranus Tilting, The Astrophysical Journal, 712, L44, 10.1088/2041-8205/712/1/L44.

Broadfoot, A. L., Herbert, F., Holberg, J. B., et al. (1986), Ultraviolet Spectrometer Observations of Uranus, Science, 233, 74, 10.1126/science.233.4759.74.

Broadfoot, A. L., Atreya, S. K., Bertaux, J. L., et al. (1989), Ultraviolet Spectrometer Observations of Neptune and Triton, Science, 246, 1459, 10.1126/science.246.4936.1459.

Brown, R. H., & Cruikshank, D. P. (1983), The Uranian satellites: Surface compositions and opposition brightness surges, Icarus, 55, 83, 10.1016/0019-1035(83)90052-0.

Buchvarova, M., & Velinov, P. (2009), Cosmic ray spectra in planetary atmospheres, Universal Heliophysical Processes, 257, 471, 10.1017/S1743921309029718.

Cao, X., & Paty, C. (2017), Diurnal and seasonal variability of Uranus's magnetosphere, Journal of Geophysical Research (Space Physics), 122, 6318, 10.1002/2017JA024063.

Cartwright, R. J., Emery, J. P., Rivkin, A. S., et al. (2015), Distribution of $CO_2$ ice on the large moons of Uranus and evidence for compositional stratification of their near-surfaces, Icarus, 257, 428, 10.1016/j.icarus.2015.05.020.

Cartwright, R. J., Emery, J. P., Pinilla-Alonso, N., et al. (2018), Red material on the large moons of Uranus: Dust from the irregular satellites?, Icarus, 314, 210, 10.1016/j.icarus.2018.06.004.

Cavalié, T., Venot, O., Selsis, F., et al. (2017), Thermochemistry and vertical mixing in the tropospheres of Uranus and Neptune: How convection inhibition can affect the derivation of deep oxygen abundances, Icarus, 291, 1, 10.1016/j.icarus.2017.03.015.

Cavalié, T., Venot, O., Miguel, Y., et al. (2020), The deep composition of Uranus and Neptune from in situ exploration and thermochemical modeling, Space Science Reviews, accepted, 10.1007/s11214-020-00677-8.

Charnoz, S., Canup, R. M., Crida, A., et al. (2018), The Origin of Planetary Ring Systems, Planetary Ring Systems. Properties, Structure, and Evolution, 517, 10.1017/9781316286791.018.

Colwell, J. E., & Esposito, L. W. (1993), Origins of the rings of Uranus and Neptune. 2. Initial conditions and ring moon populations, Journal of Geophysical Research, 98, 7387, 10.1029/93JE00329.

Colwell, J. E., & Esposito, L. W. (1992), Origins of the rings of Uranus and Neptune 1. Statistics of satellite disruptions, Journal of Geophysical Research, 97, 10227, 10.1029/92JE00788.

Conrath, B. J., Gierasch, P. J., & Ustinov, E. A. (1998), Thermal Structure and Para Hydrogen Fraction on the Outer Planets from Voyager IRIS Measurements, Icarus, 135, 501, 10.1006/icar.1998.6000.

Cowley, S. W. H. (2013), Response of Uranus' auroras to solar wind compressions at equinox, Journal of Geophysical Research (Space Physics), 118, 2897, 10.1002/jgra.50323.

Croft, S. K., & Soderblom, L. A. (1991), Geology of the Uranian satellites., Uranus, 561.

Cruikshank, D. P., Schmitt, B., Roush, T. L., et al. (2000), Water Ice on Triton, Icarus, 147, 309, 10.1006/icar.2000.6451.

de Pater, I., Romani, P. N., & Atreya, S. K. (1991), Possible microwave absorption by H 2S gas in Uranus' and Neptune's atmospheres, Icarus, 91, 220, 10.1016/0019-1035(91)90020-T.

de Pater, I., Hammel, H. B., Showalter, M. R., et al. (2007), The Dark Side of the Rings of Uranus, Science, 317, 1888, 10.1126/science.1148103.

de Pater, I., Sromovsky, L. A., Fry, P. M., et al. (2015), Record-breaking storm activity on Uranus in 2014, Icarus, 252, 121, 10.1016/j.icarus.2014.12.037.

de Pater, I., Gibbard, S. G., & Hammel, H. B. (2006), Evolution of the dusty rings of Uranus, Icarus, 180, 186, 10.1016/j.icarus.2005.08.011.

de Pater, I., Gibbard, S. G., Chiang, E., et al. (2005), The dynamic neptunian ring arcs: evidence for a gradual disappearance of Liberté and resonant jump of courage, Icarus, 174, 263, 10.1016/j.icarus.2004.10.020.

de Pater, I., Fletcher, L. N., Luszcz-Cook, S., et al. (2014), Neptune's global circulation deduced from multi-wavelength observations, Icarus, 237, 211, 10.1016/j.icarus.2014.02.030.

de Pater, I., Renner, S., Showalter, M. R., et al. (2018), The Rings of Neptune, Planetary Ring Systems. Properties, Structure, and Evolution, 112, 10.1017/9781316286791.005.

Decker, R. B., & Cheng, A. F. (1994), A model of Triton's role in Neptune's magnetosphere, Journal of Geophysical Research, 99, 19027, 10.1029/94JE01867.

Desch, M. D., Kaiser, M. L., Zarka, P., et al. (1991), Uranus as a radio source., Uranus, 894.

Dobrijevic, M., Loison, J. C., Hue, V., et al. (2020), 1D photochemical model of the ionosphere and the stratosphere of Neptune, Icarus, 335, 113375, 10.1016/j.icarus.2019.07.009.

Dodson-Robinson, S. E., & Bodenheimer, P. (2010), The formation of Uranus and Neptune in solid-rich feeding zones: Connecting chemistry and dynamics, Icarus, 207, 491, 10.1016/j.icarus.2009.11.021.

Dumas, C., Terrile, R. J., Smith, B. A., et al. (1999), Stability of Neptune's ring arcs in question, Nature, 400, 733, 10.1038/23414.

Farrell, W. M. (1992), Nonthermal radio emissions from Uranus, Planetary Radio Emissions III, 241.

Feuchtgruber, H., Lellouch, E., de Graauw, T., et al. (1997), External supply of oxygen to the atmospheres of the giant planets, Nature, 389, 159, 10.1038/38236.

Fletcher, L. N., de Pater, I., Orton, G. S., et al. (2020), Ice Giant Circulation Patterns: Implications for Atmospheric Probes, Space Science Reviews, 216, 21, 10.1007/s11214-020-00646-1.

Fletcher, L. N., de Pater, I., Orton, G. S., et al. (2014), Neptune at summer solstice: Zonal mean temperatures from ground-based observations, 2003-2007, Icarus, 231, 146, 10.1016/j.icarus.2013.11.035.

French, R. G., Nicholson, P. D., Porco, C. C., et al. (1991), Dynamics and structure of the Uranian rings., Uranus, 327.

Fressin, F., Torres, G., Charbonneau, D., et al. (2013), The False Positive Rate of Kepler and the Occurrence of Planets, The Astrophysical Journal, 766, 81, 10.1088/0004-637X/766/2/81.

Fulton, B. J., & Petigura, E. A. (2018), The California-Kepler Survey. VII. Precise Planet Radii Leveraging Gaia DR2 Reveal the Stellar Mass Dependence of the Planet Radius Gap, The Astronomical Journal, 156, 264, 10.3847/1538-3881/aae828.

Gierasch, P. J., & Conrath, B. J. (1987), Vertical temperature gradients on Uranus: Implications for layered convection, Journal of Geophysical Research, 92, 15019, 10.1029/JA092iA13p15019.

Griton, L., Pantellini, F., & Meliani, Z. (2018), Three-Dimensional Magnetohydrodynamic Simulations of the Solar Wind Interaction With a Hyperfast-Rotating Uranus, Journal of Geophysical Research (Space Physics), 123, 5394, 10.1029/2018JA025331.

Griton, L., & Pantellini, F. (2020), Magnetohydrodynamic simulations of a Uranus-at-equinox type rotating magnetosphere, Astronomy and Astrophysics, 633, A87, 10.1051/0004-6361/201936604.

Grundy, W. M., Binzel, R. P., Buratti, B. J., et al. (2016), Surface compositions across Pluto and Charon, Science, 351, aad9189, 10.1126/science.aad9189.

Grundy, W. M., Young, L. A., Spencer, J. R., et al. (2006), Distributions of $H_2O$ and $CO_2$ ices on Ariel, Umbriel, Titania, and Oberon from IRTF/SpeX observations, Icarus, 184, 543, 10.1016/j.icarus.2006.04.016.

Guillot, T. (1995), Condensation of Methane, Ammonia, and Water and the Inhibition of Convection in Giant Planets, Science, 269, 1697, 10.1126/science.7569896.

Guillot, T. (2019), Uranus and Neptune are key to understand planets with hydrogen atmospheres, arXiv e-prints, arXiv:1908.02092.

Gunell, H., Maggiolo, R., Nilsson, H., et al. (2018), Why an intrinsic magnetic field does not protect a planet

against atmospheric escape, Astronomy and Astrophysics, 614, L3, 10.1051/0004-6361/201832934.

Gurnett, D. A., Kurth, W. S., Cairns, I. H., et al. (1990), Whistlers in Neptune's magnetosphere: Evidence of atmospheric lightning, Journal of Geophysical Research, 95, 20967, 10.1029/JA095iA12p20967.

Helled, R., Bodenheimer, P., Podolak, M., et al. (2014), Giant Planet Formation, Evolution, and Internal Structure, Protostars and Planets VI, 643, 10.2458/azu_uapress_9780816531240-ch028.

Helled, R., & Bodenheimer, P. (2014), The Formation of Uranus and Neptune: Challenges and Implications for Intermediate-mass Exoplanets, The Astrophysical Journal, 789, 69, 10.1088/0004-637X/789/1/69.

Helled, R., Anderson, J. D., Podolak, M., et al. (2011), Interior Models of Uranus and Neptune, The Astrophysical Journal, 726, 15, 10.1088/0004-637X/726/1/15.

Helled, R., Anderson, J. D., & Schubert, G. (2010), Uranus and Neptune: Shape and rotation, Icarus, 210, 446, 10.1016/j.icarus.2010.06.037.

Herbert, F. (2009), Aurora and magnetic field of Uranus, Journal of Geophysical Research (Space Physics), 114, A11206, 10.1029/2009JA014394.

Herbert, F., Sandel, B. R., Yelle, R. V., et al. (1987), The upper atmosphere of Uranus: EUV occultations observed by Voyager 2, Journal of Geophysical Research, 92, 15093, 10.1029/JA092iA13p15093.

Hofstadter, M., Simon, A., Atreya, S., et al. (2019), Uranus and Neptune missions: A study in advance of the next Planetary Science Decadal Survey, Planetary and Space Science, 177, 104680, 10.1016/j.pss.2019.06.004.

Hofstadter, M. D., & Butler, B. J. (2003), Seasonal change in the deep atmosphere of Uranus, Icarus, 165, 168, 10.1016/S0019-1035(03)00174-X.

Hoogeveen, G. W., & Cloutier, P. A. (1996), The Triton-Neptune plasma interaction, Journal of Geophysical Research, 101, 19, 10.1029/95JA02761.

Hsu, H.-W., Schmidt, J., Kempf, S., et al. (2018), In situ collection of dust grains falling from Saturn's rings into its atmosphere, Science, 362, aat3185, 10.1126/science.aat3185.

Hueso, R., & Sánchez-Lavega, A. (2019), Atmospheric Dynamics and Vertical Structure of Uranus and Neptune's Weather Layers, Space Science Reviews, 215, 52, 10.1007/s11214-019-0618-6.

Hueso, R., Guillot, T., Sánchez-Lavega, A. (2020). Atmospheric dynamics and convective regimes in the non-homogeneous weather layers of the ice giants, Philosophical Transactions A. Submitted.

Hussmann, H., Sohl, F., & Spohn, T. (2006), Subsurface oceans and deep interiors of medium-sized outer planet satellites and large trans-neptunian objects, Icarus, 185, 258, 10.1016/j.icarus.2006.06.005.

Iess, L., Militzer, B., Kaspi, Y., et al. (2019), Measurement and implications of Saturn's gravity field and ring mass, Science, 364, aat2965, 10.1126/science.aat2965.

Irwin, P. G. J., Toledo, D., Garland, R., et al. (2018), Detection of hydrogen sulfide above the clouds in Uranus's atmosphere, Nature Astronomy, 2, 420, 10.1038/s41550-018-0432-1.

Jacobson, R. A. (2014), The Orbits of the Uranian Satellites and Rings, the Gravity Field of the Uranian System, and the Orientation of the Pole of Uranus, The Astronomical Journal, 148, 76, 10.1088/0004-6256/148/5/76.

Jacobson, R. A. (2009), The Orbits of the Neptunian Satellites and the Orientation of the Pole of Neptune, The Astronomical Journal, 137, 4322, 10.1088/0004-6256/137/5/4322.

Karkoschka, E., & Tomasko, M. G. (2011), The haze and methane distributions on Neptune from HST-STIS spectroscopy, Icarus, 211, 780, 10.1016/j.icarus.2010.08.013.

Kaspi, Y., Showman, A. P., Hubbard, W. B., et al. (2013), Atmospheric confinement of jet streams on Uranus and Neptune, Nature, 497, 344, 10.1038/nature12131.

Kegerreis, J. A., Eke, V. R., Gonnet, P., et al. (2019), Planetary giant impacts: convergence of high-resolution simulations using efficient spherical initial conditions and SWIFT, Monthly Notices of the Royal Astronomical Society, 487, 5029, 10.1093/mnras/stz1606.

Kempf, S., Altobelli, N., Srama, R., et al. (2018), The Age of Saturn's Rings Constrained by the Meteoroid Flux Into the System, EGU General Assembly Conference Abstracts, 10791.

Krasnopolsky, V. A., & Cruikshank, D. P. (1995), Photochemistry of Triton's atmosphere and ionosphere., Journal of Geophysical Research, 100, 21,271,286.

Lainey, V. (2008), A new dynamical model for the Uranian satellites, Planetary and Space Science, 56, 1766, 10.1016/j.pss.2008.02.015.

Lambrechts, M., Johansen, A., & Morbidelli, A. (2014), Separating gas-giant and ice-giant planets by halting pebble accretion, Astronomy and Astrophysics, 572, A35, 10.1051/0004-6361/201423814.

Lammer, H. (1995), Mass loss of N 2 molecules from Triton by magnetospheric plasma interaction, Planetary and Space Science, 43, 845, 10.1016/0032-0633(94)00214-C.

Lamy, L., Prangé, R., Hansen, K. C., et al. (2017), The aurorae of Uranus past equinox, Journal of Geophysical Research (Space Physics), 122, 3997, 10.1002/2017JA023918.

Lamy, L., Prangé, R., Hansen, K. C., et al. (2012), Earth-based detection of Uranus' aurorae, Geophysical Research Letters, 39, L07105, 10.1029/2012GL051312.

LeBeau, R. P., & Dowling, T. E. (1998), EPIC Simulations of Time-Dependent, Three-Dimensional Vortices with Application to Neptune's Great Dark SPOT, Icarus, 132, 239, 10.1006/icar.1998.5918.

Leconte, J., Selsis, F., Hersant, F., et al. (2017), Condensation-inhibited convection in hydrogen-rich atmospheres . Stability against double-diffusive processes and thermal profiles for Jupiter, Saturn, Uranus, and Neptune, Astronomy and Astrophysics, 598, A98, 10.1051/0004-6361/201629140.

Lellouch, E., de Bergh, C., Sicardy, B., et al. (2010), Detection of CO in Triton's atmosphere and the nature of surface-atmosphere interactions, Astronomy and Astrophysics, 512, L8, 10.1051/0004-6361/201014339.

Li, C., Le, T., Zhang, X., et al. (2018), A high-performance atmospheric radiation package: With applications to the radiative energy budgets of giant planets, Journal of Quantitative Spectroscopy and Radiative Transfer, 217, 353, 10.1016/j.jqsrt.2018.06.002.

Lindal, G. F., Lyons, J. R., Sweetnam, D. N., et al. (1987), The atmosphere of Uranus: Results of radio occultation measurements with Voyager 2, Journal of Geophysical Research, 92, 14987, 10.1029/JA092iA13p14987.

Lindal, G. F. (1992), The Atmosphere of Neptune: an Analysis of Radio Occultation Data Acquired with Voyager 2, The Astronomical Journal, 103, 967, 10.1086/116119.

Majeed, T., Waite, J. H., Bougher, S. W., et al. (2004), The ionospheres-thermospheres of the giant planets, Advances in Space Research, 33, 197, 10.1016/j.asr.2003.05.009.

Martin-Herrero, A., Romeo, I., & Ruiz, J. (2018), Heat flow in Triton: Implications for heat sources powering recent geologic activity, Planetary and Space Science, 160, 19, 10.1016/j.pss.2018.03.010.

Masters, A. (2014), Magnetic reconnection at Uranus' magnetopause, Journal of Geophysical Research (Space Physics), 119, 5520, 10.1002/2014JA020077.

Masters, A. (2018), A More Viscous-Like Solar Wind Interaction With All the Giant Planets, Geophysical Research Letters, 45, 7320, 10.1029/2018GL078416.

Masters, A., Achilleos, N., Agnor, C. B., et al. (2014), Neptune and Triton: Essential pieces of the Solar System puzzle, Planetary and Space Science, 104, 108, 10.1016/j.pss.2014.05.008.

Mauk, B. H., & Fox, N. J. (2010), Electron radiation belts of the solar system, Journal of Geophysical Research (Space Physics), 115, A12220, 10.1029/2010JA015660.

McKinnon, W. B., & Leith, A. C. (1995), Gas drag and the orbital evolution of a captured Triton., Icarus, 118, 392, 10.1006/icar.1995.1199.

McKinnon, W. B., Stern, S. A., Weaver, H. A., et al. (2017), Origin of the Pluto-Charon system: Constraints from the New Horizons flyby, Icarus, 287, 2, 10.1016/j.icarus.2016.11.019.

McNutt, R. L., Selesnick, R. S., & Richardson, J. D. (1987), Low-energy plasma observations in the magnetosphere of Uranus, Journal of Geophysical Research, 92, 4399, 10.1029/JA092iA05p04399.

Mejnertsen, L., Eastwood, J. P., Chittenden, J. P., et al. (2016), Global MHD simulations of Neptune's magnetosphere, Journal of Geophysical Research (Space Physics), 121, 7497, 10.1002/2015JA022272.

Melin, H., Stallard, T., Miller, S., et al. (2011), Seasonal Variability in the Ionosphere of Uranus, The Astrophysical Journal, 729, 134, 10.1088/0004-637X/729/2/134.

Melin, H., Fletcher, L. N., Stallard, T. S., et al. (2019), The H3+ ionosphere of Uranus: decades-long cooling and local-time morphology, Philosophical Transactions of the Royal Society of London Series A, 377, 20180408, 10.1098/rsta.2018.0408.

Melin, H., Stallard, T. S., Miller, S., et al. (2013), Post-equinoctial observations of the ionosphere of Uranus, Icarus, 223, 741, 10.1016/j.icarus.2013.01.012.

Merrill, R. T., & McFadden, P. L. (1999), Geomagnetic polarity transitions, Reviews of Geophysics, 37, 201, 10.1029/1998RG900004.

Millot, M., Coppari, F., Rygg, J. R., et al. (2019), Nanosecond X-ray diffraction of shock-compressed superionic water ice, Nature, 569, 251, 10.1038/s41586-019-1114-6.

Molter, E. M., de Pater, I., Roman, M. T., et al. (2019), Thermal Emission from the Uranian Ring System, The Astronomical Journal, 158, 47, 10.3847/1538-3881/ab258c.

Moreno, R., Marten, A., & Lellouch, E. (2009), Search for PH3 in the Atmospheres of Uranus and Neptune at Millimeter Wavelength, AAS/Division for Planetary Sciences Meeting Abstracts #41, 28.02.

Morley, C. V., Fortney, J. J., Marley, M. S., et al. (2012), Neglected Clouds in T and Y Dwarf Atmospheres, The Astrophysical Journal, 756, 172, 10.1088/0004-637X/756/2/172.

Moses, J. I., & Poppe, A. R. (2017), Dust ablation on the giant planets: Consequences for stratospheric photochemistry, Icarus, 297, 33, 10.1016/j.icarus.2017.06.002.

Moses, J. I., Fletcher, L. N., Greathouse, T. K., et al. (2018), Seasonal stratospheric photochemistry on Uranus and Neptune, Icarus, 307, 124, 10.1016/j.icarus.2018.02.004.

Mousis, O., Atkinson, D. H., Cavalié, T., et al. (2018), Scientific rationale for Uranus and Neptune in situ explorations, Planetary and Space Science, 155, 12, 10.1016/j.pss.2017.10.005.

Namouni, F., & Porco, C. (2002), The confinement of Neptune's ring arcs by the moon Galatea, Nature, 417, 45, 10.1038/417045a.

Nettelmann, N., Helled, R., Fortney, J. J., et al. (2013), New indication for a dichotomy in the interior structure of Uranus and Neptune from the application of modified shape and rotation data, Planetary and Space Science, 77, 143, 10.1016/j.pss.2012.06.019.

Neubauer, F. M. (1990), Satellite plasma interactions, Advances in Space Research, 10, 25, 10.1016/0273-1177(90)90083-C.

Nicholson, P. D., De Pater, I., French, R. G., et al. (2018), The Rings of Uranus, Planetary Ring Systems. Properties, Structure, and Evolution, 93, 10.1017/9781316286791.004.

Nimmo, F., & Spencer, J. R. (2015), Powering Triton's recent geological activity by obliquity tides: Implications for Pluto geology, Icarus, 246, 2, 10.1016/j.icarus.2014.01.044.

Orton, G. S., Fletcher, L. N., Moses, J. I., et al. (2014), Mid-infrared spectroscopy of Uranus from the Spitzer Infrared Spectrometer: 1. Determination of the mean temperature structure of the upper troposphere and stratosphere, Icarus, 243, 494, 10.1016/j.icarus.2014.07.010.

Pearl, J. C., Conrath, B. J., Hanel, R. A., et al. (1990), The albedo, effective temperature, and energy balance of Uranus, as determined from Voyager IRIS data, Icarus, 84, 12, 10.1016/0019-1035(90)90155-3.

Pearl, J. C., & Conrath, B. J. (1991), The albedo, effective temperature, and energy balance of Neptune, as determined from Voyager data, Journal of Geophysical Research, 96, 18921, 10.1029/91JA01087.

Penny, M. T., Gaudi, B. S., Kerins, E., et al. (2019), Predictions of the WFIRST Microlensing Survey. I. Bound Planet Detection Rates, The Astrophysical Journal Supplement Series, 241, 3, 10.3847/1538-4365/aafb69.

Petigura, E. A., Howard, A. W., Marcy, G. W., et al. (2017), The California-Kepler Survey. I. High-resolution Spectroscopy of 1305 Stars Hosting Kepler Transiting Planets, The Astronomical Journal, 154, 107, 10.3847/1538-3881/aa80de.

Plainaki, C., Lilensten, J., Radioti, A., et al. (2016), Planetary space weather: scientific aspects and future perspectives, Journal of Space Weather and Space Climate, 6, A31, 10.1051/swsc/2016024.

Plescia, J. B. (1987a), Cratering history of the Uranian satellites: Umbriel, Titania, and Oberon, Journal of Geophysical Research, 92, 14918, 10.1029/JA092iA13p14918.

Plescia, J. B. (1987b), Geological terrains and crater frequencies on Ariel, Nature, 327, 201, 10.1038/327201a0.

Podolak, M., & Helled, R. (2012), What Do We Really Know about Uranus and Neptune?, The Astrophysical Journal, 759, L32, 10.1088/2041-8205/759/2/L32.

Podolak, M., Weizman, A., & Marley, M. (1995), Comparative models of Uranus and Neptune, Planetary and Space Science, 43, 1517, 10.1016/0032-0633(95)00061-5.

Pollack, J. B., Hubickyj, O., Bodenheimer, P., et al. (1996), Formation of the Giant Planets by Concurrent Accretion of Solids and Gas, Icarus, 124, 62, 10.1006/icar.1996.0190.

Postberg, F., Kempf, S., Schmidt, J., et al. (2009), Sodium salts in E-ring ice grains from an ocean below the surface of Enceladus, Nature, 459, 1098, 10.1038/nature08046.

Prockter, L. M., Nimmo, F., & Pappalardo, R. T. (2005), A shear heating origin for ridges on Triton, Geophysical Research Letters, 32, L14202, 10.1029/2005GL022832.

Quirico, E., Douté, S., Schmitt, B., et al. (1999), Composition, Physical State, and Distribution of Ices at the Surface of Triton, Icarus, 139, 159, 10.1006/icar.1999.6111.

Rages, K., Pollack, J. B., Tomasko, M. G., et al. (1991), Properties of scatterers in the troposphere and lower stratosphere of Uranus based on Voyager imaging data, Icarus, 89, 359, 10.1016/0019-1035(91)90183-T.

Redmer, R., Mattsson, T. R., Nettelmann, N., et al. (2011), The phase diagram of water and the magnetic fields of Uranus and Neptune, Icarus, 211, 798, 10.1016/j.icarus.2010.08.008.

Reinhardt, C., Chau, A., Stadel, J., et al. (2020), Bifurcation in the history of Uranus and Neptune: the role of giant impacts, Monthly Notices of the Royal Astronomical Society, 492, 5336, 10.1093/mnras/stz3271.

Richardson, J. D., & McNutt, R. L. (1990), Low-energy plasma in Neptune's magnetosphere, Geophysical Research Letters, 17, 1689, 10.1029/GL017i010p01689.

Roman, M. T., Fletcher, L. N., Orton, G. S., et al. (2020), Uranus in Northern Midspring: Persistent Atmospheric Temperatures and Circulations Inferred from Thermal Imaging, The Astronomical Journal, 159, 45, 10.3847/1538-3881/ab5dc7.

Rymer, A., Mandt, K., Hurley, D., et al. (2019), Solar System Ice Giants: Exoplanets in our Backyard., Bulletin of the American Astronomical Society, 51, 176.

Safronov, V. S. (1966), Sizes of the largest bodies falling onto the planets during their formation, Soviet Astronomy, 9, 987.

Sánchez-Lavega, A., Sromovsky, L. A., et al. (2018), Gas Giants. In: Galperin, B., Read, P.L., (eds) Zonal Jets Phenomenology, Genesis and Physics. Cambridge (doi: 10.1017/9781107358225).

Scarf, F. L., Gurnett, D. A., Kurth, W. S., et al. (1987), Voyager 2 plasma wave observations at Uranus, Advances in Space Research, 7, 253, 10.1016/0273-1177(87)90226-2.

Selesnick, R. S. (1988), Magnetospheric convection in the nondipolar magnetic field of Uranus, Journal of Geophysical Research, 93, 9607, 10.1029/JA093iA09p09607.

Showalter, M. R., & Lissauer, J. J. (2006), The Second Ring-Moon System of Uranus: Discovery and Dynamics, Science, 311, 973, 10.1126/science.1122882.

Simon, A. A., Fletcher, L. N., Arridge, C., et al. (2020), A Review of the in Situ Probe Designs from Recent Ice Giant Mission Concept Studies, Space Science Reviews, 216, 17, 10.1007/s11214-020-0639-1.

Simon, A. A., Wong, M. H., & Hsu, A. I. (2019), Formation of a New Great Dark Spot on Neptune in 2018, Geophysical Research Letters, 46, 3108, 10.1029/2019GL081961.

Simon, A. A., Stern, S. A., & Hofstadter, M. (2018), Outer Solar System Exploration: A Compelling and Unified Dual Mission Decadal Strategy for Exploring Uranus, Neptune, Triton, Dwarf Planets, and Small KBOs and Centaurs, arXiv e-prints, arXiv:1807.08769.

Simon, A. A., Rowe, J. F., Gaulme, P., et al. (2016), Neptune's Dynamic Atmosphere from Kepler K2 Observations: Implications for Brown Dwarf Light Curve Analyses, The Astrophysical Journal, 817, 162, 10.3847/0004-637X/817/2/162.

Slattery, W. L., Benz, W., & Cameron, A. G. W. (1992), Giant impacts on a primitive Uranus, Icarus, 99, 167, 10.1016/0019-1035(92)90180-F.

Smith, B. A., Soderblom, L. A., Banfield, D., et al. (1989), Voyager 2 at Neptune: Imaging Science Results, Science, 246, 1422, 10.1126/science.246.4936.1422.

Smith, M. D., & Gierasch, P. J. (1995), Convection in the outer planet atmospheres including ortho-para hydrogen conversion., Icarus, 116, 159, 10.1006/icar.1995.1118.

Soderblom, L. A., Kieffer, S. W., Becker, T. L., et al. (1990), Triton's Geyser-Like Plumes: Discovery and Basic Characterization, Science, 250, 410, 10.1126/science.250.4979.410.

Soderlund, K. M., Heimpel, M. H., King, E. M., et al. (2013), Turbulent models of ice giant internal dynamics: Dynamos, heat transfer, and zonal flows, Icarus, 224, 97, 10.1016/j.icarus.2013.02.014.

Sromovsky, L. A., de Pater, I., Fry, P. M., et al. (2015), High S/N Keck and Gemini AO imaging of Uranus during 2012-2014: New cloud patterns, increasing activity, and improved wind measurements, Icarus, 258, 192, 10.1016/j.icarus.2015.05.029.

Sromovsky, L. A., Karkoschka, E., Fry, P. M., et al. (2014), Methane depletion in both polar regions of Uranus inferred from HST/STIS and Keck/NIRC2 observations, Icarus, 238, 137, 10.1016/j.icarus.2014.05.016.

Stanley, S., & Bloxham, J. (2004), Convective-region geometry as the cause of Uranus' and Neptune's unusual magnetic fields, Nature, 428, 151, 10.1038/nature02376.

Stanley, S., & Bloxham, J. (2006), Numerical dynamo models of Uranus' and Neptune's magnetic fields, Icarus, 184, 556, 10.1016/j.icarus.2006.05.005.

Stauffer, J., Marley, M. S., Gizis, J. E., et al. (2016), Spitzer Space Telescope Mid-IR Light Curves of Neptune, The Astronomical Journal, 152, 142, 10.3847/0004-6256/152/5/142.

Stern, S. A., & McKinnon, W. B. (2000), Triton's Surface Age and Impactor Population Revisited in Light of Kuiper Belt Fluxes: Evidence for Small Kuiper Belt Objects and Recent Geological Activity, The Astronomical Journal, 119, 945, 10.1086/301207.

Stevenson, D. J. (1986), The Uranus-Neptune Dichotomy: the Role of Giant Impacts, Lunar and Planetary Science Conference, 1011.

Stoker, C. R., & Toon, O. B. (1989), Moist convection on Neptune, Geophysical Research Letters, 16, 929, 10.1029/GL016i008p00929.

Stone, E. C., Cooper, J. F., Cummings, A. C., et al. (1986), Energetic Charged Particles in the Uranian Magnetosphere, Science, 233, 93, 10.1126/science.233.4759.93.

Stratman, P. W., Showman, A. P., Dowling, T. E., et al. (2001), EPIC Simulations of Bright Companions to Neptune's Great Dark Spots, Icarus, 151, 275, 10.1006/icar.2001.6603.

Strobel, D. F., Cheng, A. F., Summers, M. E. (1990), Magnetospheric interaction with Triton's ionosphere, Geophysical Research Letters, 17, 1661, 10.1029/GL017i010p01661.

Stryk, T., & Stooke, P. J. (2008), Voyager 2 Images of Uranian Satellites: Reprocessing and New Interpretations, Lunar and Planetary Science Conference, 1362.

Sun, Z.-P., Schubert, G., & Stoker, C. R. (1991), Thermal and humidity winds in outer planet atmospheres, Icarus, 91, 154, 10.1016/0019-1035(91)90134-F.

Tegler, S. C., Grundy, W. M., Olkin, C. B., et al. (2012), Ice Mineralogy across and into the Surfaces of Pluto, Triton, and Eris, The Astrophysical Journal, 751, 76, 10.1088/0004-637X/751/1/76.

Tegler, S. C., Stufflebeam, T. D., Grundy, W. M., et al. (2019), A New Two-molecule Combination Band as a Diagnostic of Carbon Monoxide Diluted in Nitrogen Ice on Triton, The Astronomical Journal, 158, 17, 10.3847/1538-3881/ab199f.

Thompson, W. R., & Sagan, C. (1990), Color and chemistry on Triton, Science, 250, 415, 10.1126/science.11538073.

Tiscareno, M. S., Hedman, M. M., Burns, J. A., et al. (2013), Compositions and Origins of Outer Planet Systems: Insights from the Roche Critical Density, The Astrophysical Journal, 765, L28, 10.1088/2041-8205/765/2/L28.

Tittemore, W. C., & Wisdom, J. (1990), Tidal evolution of the Uranian satellites   III. Evolution through the Miranda-Umbriel 3:1, Miranda-Ariel 5:3, and Ariel-Umbriel 2:1 mean-motion commensurabilities, Icarus, 85, 394, 10.1016/0019-1035(90)90125-S.

Tosi, F., Turrini, D., Coradini, A., et al. (2010), Probing the origin of the dark material on Iapetus, Monthly Notices of the Royal Astronomical Society, 403, 1113, 10.1111/j.1365-2966.2010.16044.x.

Turrini, D., Politi, R., Peron, R., et al. (2014), The comparative exploration of the ice giant planets with twin spacecraft: Unveiling the history of our Solar System, Planetary and Space Science, 104, 93, 10.1016/j.pss.2014.09.005.

Venturini, J., & Helled, R. (2017), The Formation of Mini-Neptunes, The Astrophysical Journal, 848, 95, 10.3847/1538-4357/aa8cd0.

Waite, J. H., Perryman, R. S., Perry, M. E., et al. (2018), Chemical interactions between Saturn's atmosphere and its rings, Science, 362, aat2382, 10.1126/science.aat2382.

Wakeford, H. R., Visscher, C., Lewis, N. K., et al. (2017), High-temperature condensate clouds in super-hot Jupiter atmospheres, Monthly Notices of the Royal Astronomical Society, 464, 4247, 10.1093/mnras/stw2639.

Wang, C., & Richardson, J. D. (2004), Interplanetary coronal mass ejections observed by Voyager 2 between 1 and 30 AU, Journal of Geophysical Research (Space Physics), 109, A06104, 10.1029/2004JA010379.

Wei, Y., Pu, Z., Zong, Q., et al. (2014), Oxygen escape from the Earth during geomagnetic reversals: Implications to mass extinction, Earth and Planetary Science Letters, 394, 94, 10.1016/j.epsl.2014.03.018.

Witasse, O., Sánchez-Cano, B., Mays, M. L., et al. (2017), Interplanetary coronal mass ejection observed at STEREO-A, Mars, comet 67P/Churyumov-Gerasimenko, Saturn, and New Horizons en route to Pluto: Comparison of its Forbush decreases at 1.4, 3.1, and 9.9 AU, Journal of Geophysical Research (Space Physics), 122, 7865, 10.1002/2017JA023884.

Wong, M. H., Tollefson, J., Hsu, A. I., et al. (2018), A New Dark Vortex on Neptune, The Astronomical Journal, 155, 117, 10.3847/1538-3881/aaa6d6.

Zarka, P., & Pedersen, B. M. (1986), Radio detection of uranian lightning by Voyager 2, Nature, 323, 605, 10.1038/323605a0.

Zarka, P., Pedersen, B. M., Lecacheux, A., et al. (1995), Radio emissions from Neptune., Neptune and Triton, 341.

Zhang, Z., Hayes, A. G., Janssen, M. A., et al. (2017), Exposure age of Saturn's A and B rings, and the Cassini Division as suggested by their non-icy material content, Icarus, 294, 14, 10.1016/j.icarus.2017.04.008.